\pdfoutput=1

\documentclass[11pt,twoside,a4paper,cmspaper,final,collab]{cms-tdr}
\def\svnVersion{b96bc3e-D}\def\svnDate{2020/03/19}\def\cmsCernNoTag{CERN-EP-2019-141}\def\cmsCernDate{\today}\def\cmsMessage{"Published in the Journal of High Energy Physics as \href{http://dx.doi.org/10.1007/JHEP03(2020)025}{\doi{10.1007/JHEP03(2020)025}.}"}

\begin{document}\cmsNoteHeader{EXO-18-011}

\hyphenation{had-ron-i-za-tion}
\hyphenation{cal-or-i-me-ter}
\hyphenation{de-vices}
\RCS$HeadURL$
\RCS$Id$

\newlength\cmsFigWidth
\ifthenelse{\boolean{cms@external}}{\setlength\cmsFigWidth{0.49\textwidth}}{\setlength\cmsFigWidth{0.65\textwidth}}
\ifthenelse{\boolean{cms@external}}{\providecommand{\cmsLeft}{upper\xspace}}{\providecommand{\cmsLeft}{left\xspace}}
\ifthenelse{\boolean{cms@external}}{\providecommand{\cmsRight}{lower\xspace}}{\providecommand{\cmsRight}{right\xspace}}

\providecommand{\NA}{\ensuremath{\text{---}}}

\providecommand{\cmsTable}[1]{\resizebox{\textwidth}{!}{#1}}
\newlength\cmsTabSkip\setlength{\cmsTabSkip}{1ex}

\newcommand{\ttH}{\ensuremath{\ttbar\Ph}}
\newcommand{\HZZfl}{\ensuremath{\Ph\to \PZ\PZ \to 4\ell}\xspace}
\newcommand{\hzzfl}{\ensuremath{\Ph\to \PZ\PZ \to 4\ell}\xspace}
\newcommand{\mllll}{\ensuremath{m_{4\ell}}\xspace}
\newcommand{\ggZZ}{\ensuremath{\Pg\Pg \to \PZ\PZ}\xspace}
\newcommand{\met}{\ptmiss}
\newcommand{\pTmissmp}{\ensuremath{{p_{\mathrm{T},\text{mp}}^{\text{miss}}}}\xspace}
\newcommand{\pTmissproj}{\ensuremath{{p_{\mathrm{T},\text{proj}}^{\text{miss}}}}\xspace}
\newcommand{\wj}{\ensuremath{\PW+\text{jets}}\xspace}
\newcommand{\WW}{\ensuremath{\PW\PW}\xspace}
\newcommand{\ZZ}{\ensuremath{\PZ\PZ}\xspace}
\newcommand{\tautau}{\ensuremath{\tau\tau}\xspace}
\newcommand{\tauetauh}{\ensuremath{\Pe\tauh}}
\newcommand{\taumutauh}{\ensuremath{\Pgm\tauh}}
\newcommand{\tauhtauh}{\ensuremath{\tauh\tauh}}
\newcommand{\mmmm}{\ensuremath{4\PGm}}
\newcommand{\eemm}{\ensuremath{2\Pe2\PGm}}
\newcommand{\tw}{\ensuremath{\PQt\PW}}
\newcommand{\mH}{\ensuremath{m_{\Ph}}\xspace}
\newcommand{\mZp}{\ensuremath{m_\zp}\xspace}
\newcommand{\mzp}{\ensuremath{m_\zp}\xspace}
\newcommand{\mChi}{\ensuremath{m_\chi}\xspace}
\newcommand{\mdm}{\ensuremath{m_\chi}\xspace}
\newcommand{\mchi}{\ensuremath{m_\chi}\xspace}
\newcommand{\Lep}{\ensuremath{\ell}}
\newcommand{\mll}{\ensuremath{m_{\Lep\Lep}}\xspace}
\newcommand{\mA}{\ensuremath{m_{\PA}}\xspace}
\newcommand{\ptll}{\ensuremath{\pt^{\Lep \Lep}}}
\newcommand{\ptla}{\ensuremath{\pt^{\Lep _1}}}
\newcommand{\ptlb}{\ensuremath{\pt^{\Lep _2}}}
\newcommand{\delphill}{\ensuremath{\Delta\phi_{\Lep\Lep}}}
\newcommand{\delphilmetl}{\ensuremath{\Delta\phi_{\met\Lep_1}}}
\newcommand{\delphilmetll}{\ensuremath{\Delta\phi_{\met\Lep_{i}}}}
\newcommand{\delphilmett}{\ensuremath{\Delta\phi_{\met\Lep_2}}}
\newcommand{\delRll}{\ensuremath{\Delta R_{\Lep\Lep}}\xspace}
\newcommand{\mTwl}{\ensuremath{\mT^{\PW_1}}}
\newcommand{\mTwll}{\ensuremath{\mT^{\PW_{i}}}}
\newcommand{\mTwt}{\ensuremath{\mT^{\PW_2}}}
\newcommand{\tkmet}{\ensuremath{\mathrm{tracker}~\met}}
\newcommand{\mTH}{\ensuremath{\mT^{\Ph}}}
\newcommand{\dytt}{\ensuremath{\PZ/\gamma^*\to\tau^+\tau^-}\xspace}
\newcommand{\ggww}{\ensuremath{\Pg\Pg \to \PW^+\PW^-}\xspace}
\newcommand{\qqww}{\ensuremath{\PQq\PQq \to\PW^+\PW^-}\xspace}
\newcommand{\wz}{\ensuremath{\PW\PZ}\xspace}
\newcommand{\wgs}{\ensuremath{\PW\PGg^{*}}\xspace}
\newcommand{\emu}{\ensuremath{\Pe\PGm}\xspace}
\newcommand{\hbb}{\ensuremath{ \Ph \to \PQb \PQb}\xspace}
\newcommand{\hbbp}{\ensuremath{ \Ph (\to \PQb \PQb) }\xspace}
\newcommand{\hgg}{\ensuremath{ \Ph \to \gamma\gamma}\xspace}
\newcommand{\hggp}{\ensuremath{ \Ph (\to \gamma\gamma)}\xspace}
\newcommand{\htt}{\ensuremath{ \Ph \to \Pgt\Pgt}\xspace}
\newcommand{\http}{\ensuremath{ \Ph (\to \Pgt\Pgt)}\xspace}
\newcommand{\hww}{\ensuremath{ \Ph \to \WW}\xspace}
\newcommand{\hwwp}{\ensuremath{ \Ph (\to \WW)}\xspace}
\newcommand{\hzz}{\ensuremath{ \Ph \to \PZ\PZ}\xspace}
\newcommand{\hzzp}{\ensuremath{ \Ph (\to \PZ\PZ)}\xspace}
\newcommand{\ptww}{\ensuremath{\pt^{\PW\PW}}\xspace}
\newcommand{\mMed}{\ensuremath{m_{\mathrm{med}}}}
\newcommand{\PZp}{\PZpr}
\newcommand{\PA}{\PSA}
\newcommand{\PhB}{\ensuremath{\Ph_{\PQb}}\xspace}
\newcommand{\SigSI}{\ensuremath{\sigma^{\mathrm{SI}}}\xspace}
\providecommand{\NA}{\ensuremath{\text{---}}}
\ifthenelse{\boolean{cms@external}}{\providecommand{\CL}{C.L.\xspace}}{\providecommand{\CL}{CL\xspace}}
\newcommand{\gq}{\ensuremath{g_{\cPq}}}
\newcommand{\gZp}{\ensuremath{g_{\PZpr}}}
\newcommand{\gChi}{\ensuremath{g_{\chi}}}

\cmsNoteHeader{EXO-18-011}
\title{Search for dark matter particles produced in association with a Higgs boson in proton-proton collisions at $\sqrt{s} = 13\TeV$}

\date{\today}

\abstract{
A search for dark matter (DM) particles is performed using events with a Higgs boson candidate and  large missing transverse momentum. The analysis is based on proton-proton collision data at a center-of-mass energy of 13\TeV collected by the CMS experiment at the LHC in 2016, corresponding to an integrated luminosity of 35.9\fbinv. The search is performed in five Higgs boson decay channels: $\Ph \to \bbbar$, $\PGg\PGg$, $\Pgt^{+}\Pgt^{-}$, $\PW^{+}\PW^{-}$, and $\PZ\PZ$. The results from the individual channels are combined to maximize the sensitivity of the analysis.  No significant excess over the expected standard model background is observed in any of the five channels or in their combination. Limits are set on DM production in the context of two simplified models.
The results are also interpreted in terms of a spin-independent DM-nucleon scattering cross section and compared to those from direct-detection DM experiments.  This is the first search for DM particles produced in association with a Higgs boson decaying to a pair of $\PW$ or $\PZ$ bosons, and the first statistical combination based on five Higgs boson decay channels.}

\hypersetup{
pdfauthor={CMS Collaboration},
pdftitle={Search for dark matter particles produced in association with a Higgs boson in proton-proton collisions at sqrt(s) = 13 TeV},
pdfsubject={CMS},
pdfkeywords={CMS, physics, dark matter, mono-Higgs, combination}}

\maketitle

\section{Introduction}\label{sec:introduction}

A host of astrophysical and cosmological observations confirm~\cite{Gaitskell:2004gd, Trimble:1987ee, Porter:2011nv, ref:DMevi} that dark matter (DM) exists and makes up  26.4\% of the total energy density of the universe~\cite{Ade:2015xua}. However, all of the existing evidence for DM is based only on its gravitational interaction. Whether DM interacts with standard model (SM) particles in any other way remains an open question.  There are a number of beyond-the-SM theories suggesting a particle nature of DM~\cite{ref:DMNPBSM}. Several types of particle candidates for DM are proposed in these models, all compatible with the observed relic density of DM in the universe~\cite{ref:relic}. A favored hypothesis is that the bulk of DM is in the form of stable, electrically neutral, weakly interacting massive particles (WIMPs)~\cite{ref:WIMPs}, with masses in a range between a few\GeV and a few\TeV, thus opening the possibility of DM production at high-energy colliders~\cite{Beltran:2010ww}.

Traditionally, searches for DM at colliders involve a pair of WIMPs that recoil against a visible SM particle or a set of SM particles. Because of the lack of electric charge and the small interaction cross section, WIMPs do not leave a directly detectable signal, but in a hadron collider experiment their presence can be inferred via an imbalance in the total momentum in the plane transverse to the colliding beams (\ptvecmiss), as reconstructed in the detector. This scenario gives rise to a potential signature where a set of SM particles, \PX, are produced recoiling against the DM particles, represented by the \ptvecmiss (the ``mono-$\PX$" signature). Recent searches at the CERN LHC considered \PX to be a hadronic jet~\cite{ref:monojet,ref:monojet1}, heavy-flavor quarks (bottom and top)~\cite{ref:monob2,ref:monob}, a photon~\cite{ref:monog,ref:monog1}, or a \PW\ or \PZ boson~\cite{ref:monoW,Sirunyan:2017qfc,ref:monojet1,ref:monoW1}.

The discovery of an SM-like Higgs boson~\cite{ATLAS-Higgs,CMS-Higgs1,CMS-Higgs2} extended the possibility of probing DM at colliders, complementing other mono-\PX searches. In this paper we designate the state observed at 125\GeV by the symbol \Ph, since in the context of the theoretical models considered below, it does not correspond to the SM Higgs boson. Here, we present a search for the pair production of DM particles in association with a Higgs boson resulting in the final state  $\Ph+\ptmiss$~\cite{ref:monoH,ref:monoH4}, referred to as the ``mono-Higgs".  While in a typical mono-\PX search, the \PX particle is emitted as initial-state radiation, this process is strongly suppressed for the case of the Higgs boson because of the smallness of both the Higgs boson Yukawa couplings to light quarks and its loop-suppressed coupling to gluons. Thus, the mono-Higgs production can be either a result of final-state radiation of DM particles,
or of a beyond-the-SM interaction of DM particles with the Higgs boson, typically via a mediator particle. A number of searches have been carried out by the ATLAS and CMS Collaborations looking for the mono-Higgs signature in several Higgs boson decay channels, at center-of-mass energies  of 8 and 13\TeV~\cite{Aad:2015yga,Aad:2015dva,Aaboud:2016obm,CMS_MonoH_13TeV,Aaboud:2017uak,Aaboud:2017yqz,b2g:monohbb2016,exo:monohbb2016,exo:monohggtt2016}. So far, none of these searches has observed a significant excess of events over the SM expectations.

In this paper, we describe the first search for mono-Higgs production in the $\PW^{+}\PW^{-}$ and $\PZ\PZ$ Higgs boson decay channels, as well as the combination of these searches with the previously published results in the $\PQb\PAQb$~\cite{b2g:monohbb2016,exo:monohbb2016}, $\PGg\PGg$~\cite{exo:monohggtt2016}, and $\Pgt^{+}\Pgt^{-}$~\cite{exo:monohggtt2016} channels. (Hereafter, for simplicity we refer to $\PQb\PAQb$, $\Pgt^{+}\Pgt^{-}$ and $\PW^{+}\PW^{-}$ as $\PQb\PQb$, $\Pgt\Pgt$ and $\PW\PW$, respectively.) All the analyses are based on a data sample of proton-proton ($\Pp\Pp$) collisions at $\sqrt{s} = 13\TeV$ collected in 2016 and corresponding to an integrated luminosity of 35.9\fbinv.

\begin{figure}[!h]
  \centering
  \includegraphics[width=0.475\textwidth]{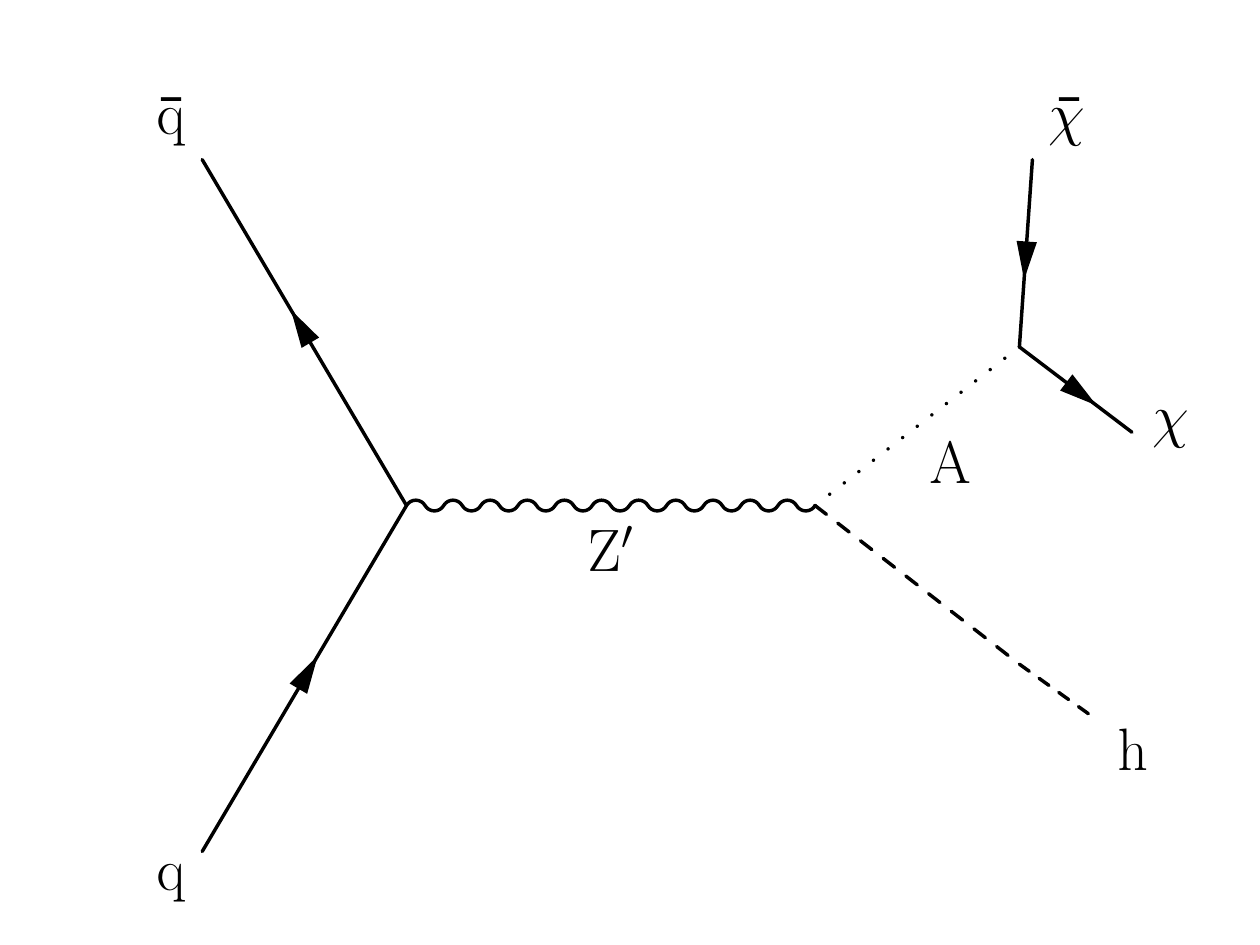}
  \includegraphics[width=0.475\textwidth]{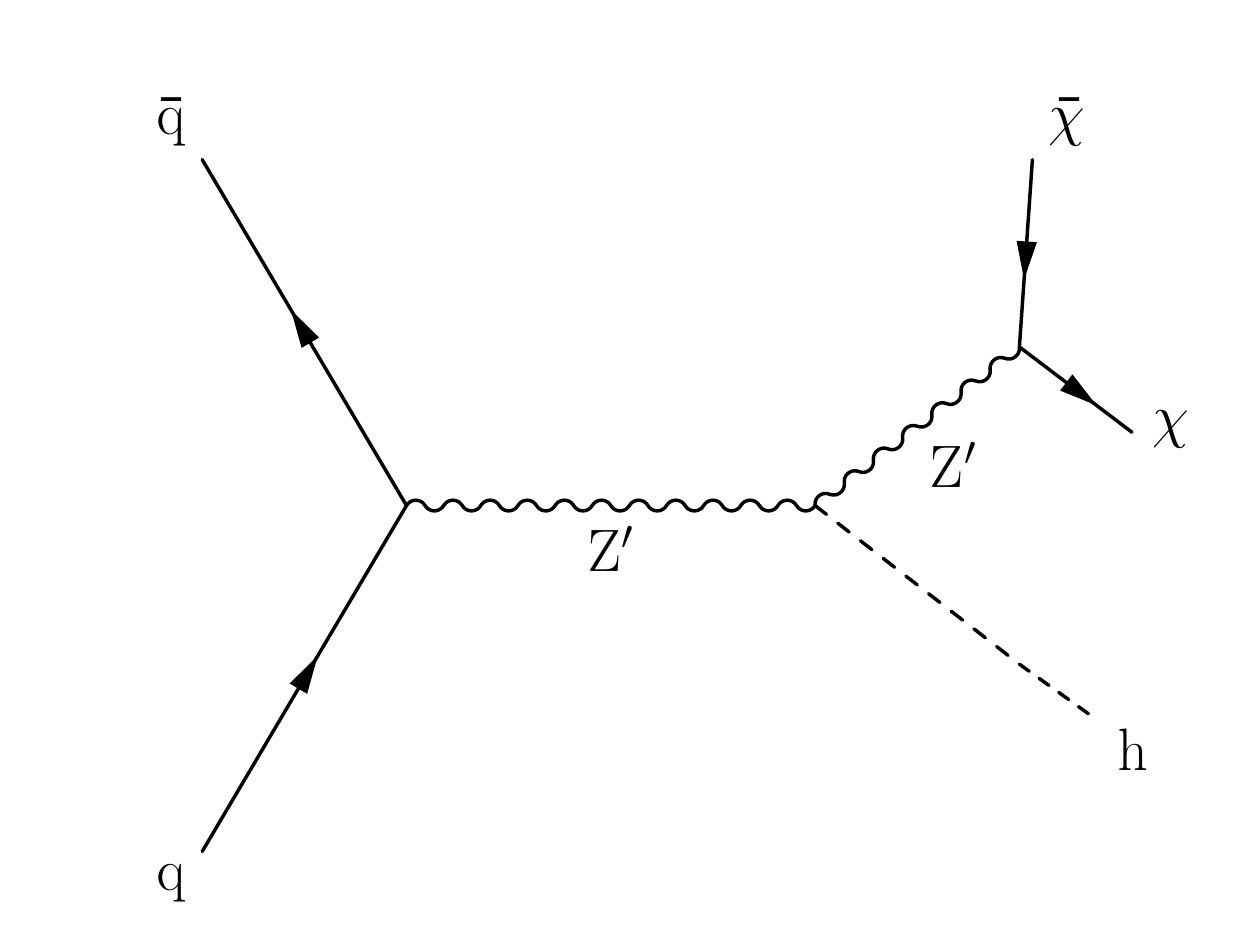}
  \caption{  Representative Feynman diagrams for the two benchmark signal models considered in this paper: the \PZp-2HDM (left) and the baryonic \PZp model (right).}
  \label{fig:models}
\end{figure}

Two simplified models of DM production recommended by the ATLAS-CMS Dark Matter Forum~\cite{Abercrombie:2015wmb} are investigated. Figure \ref{fig:models} shows representative tree-level Feynman diagrams corresponding to these two models. The diagram on the left describes a type-II two Higgs doublet model (2HDM)~\cite{Branco:2011iw,Craig:2013hca} further extended by a $U(1)_{\PZp}$ group and
referred to as the \PZp-2HDM~\cite{Z'-2HDM}. In this model, the \PZp boson is produced via a quark-antiquark interaction and then decays into a Higgs boson and a pseudoscalar mediator  \PA, which in turn can decay to a pair of Dirac fermion DM particles $\PGc$. The diagram on the right shows the production mechanism in the baryonic \PZp model~\cite{ref:monoH}, where \PZp is a vector boson corresponding to a new baryon number $U(1)_{\mathrm{B}}$ symmetry. The \PZp boson acts as a DM mediator and can radiate a Higgs boson before decaying to a pair of DM particles. A baryonic Higgs boson \PhB is introduced to spontaneously break the new symmetry and to generate the \PZp boson mass via a coupling that is dependent on the \PhB vacuum expectation value.  The \PZp boson couplings to quarks and the DM particles are proportional to the $U(1)_{\mathrm{B}}$ gauge couplings. A mixing between the \PhB and \Ph states allows the \PZp boson to radiate h, resulting in a mono-Higgs signature.

In the \PZp-2HDM, the predicted DM production cross section depends on number of parameters. However, if the mediator \PA is produced on-shell, the kinematic distributions of the final-state particles depend only on the \PZp and \PA boson masses, \mZp and \mA.
In this paper, a scan in \mZp between 450 and 4000\GeV and in \mA between 300 and 1000\GeV is performed. The values of \mA below 300\GeV have been already excluded by the existing constraints on flavor changing neutral currents in the
$\PQb\to \PQs\PGg$ transitions~\cite{Branco:2011iw}, and hence are not considered in the analysis.
The masses of the 2HDM heavy Higgs boson and the charged Higgs boson are both fixed to the \mA mass.
The ratio of the vacuum expectation values of the two Higgs doublets, $\tan{\beta}$, is varied from 0.4 to 10.
The DM particle mass is fixed to 100\GeV, the $\PA$-DM coupling strength $g_{\PGc}$ is fixed to 1,
and the \PZp coupling strength to quarks \gZp~is fixed to 0.8. The branching fraction of the decay of \PA to DM particles $\mathcal{B}$($\PA \to \PGc\bar\PGc$) decreases as the mass of the DM candidate (\mchi) increases, for the range of \mA considered in this analysis. However, since the relative decrease in
$\mathcal{B}$($\PA \to \PGc\bar\PGc$) is less than 7\% as \mchi increases from 1 to 100\GeV, the results shown in this paper for $\mchi = 100\GeV$ are also applicable to lighter DM particles.

The results are expressed in terms of the product of the signal production cross section and branching fraction $\mathcal{B}$($\PA \to \PGc\bar\PGc$), where $\mathcal{B}$($\PA \to \PGc\bar\PGc$)
is $\approx$$100\%$ for $\mA = 300\GeV$ and decreases for \mA greater than twice the mass of the top quark, where the competing decay $\PA \to \ttbar$ becomes kinematically accessible.
The contribution to the mono-Higgs signal from another process possible in the model, $\PZp \to \PZ(\to \PGn\PAGn)+\Ph$, is not considered in this analysis.
Further details on the choice of the model parameters are given in Refs.~\cite{CMS_MonoH_13TeV,presentDM}. We note that for the chosen set of parameters, the values of \mZp within our sensitivity reach have been recently excluded by the ATLAS and CMS searches for dijet resonances at $\sqrt{s} = 13\TeV$~\cite{Aaboud:2017yvp,Sirunyan:2017nvi,Aaboud:2018fzt,Sirunyan:2018xlo}. Nevertheless, we keep this benchmark, specifically developed for the LHC Run-2 searches~\cite{Abercrombie:2015wmb}, to allow a direct comparison with the results of other mono-Higgs searches. Given that the kinematic distributions of the final states depend only very weakly on the value of the $g_{\PZp}$ coupling, our results can be reinterpreted for lower $g_{\PZp}$ values, where the interplay between the mono-Higgs and the dijet analysis sensitivities changes.

\begin{figure}[htbp]
  \centering
    \includegraphics[width=0.475\textwidth]{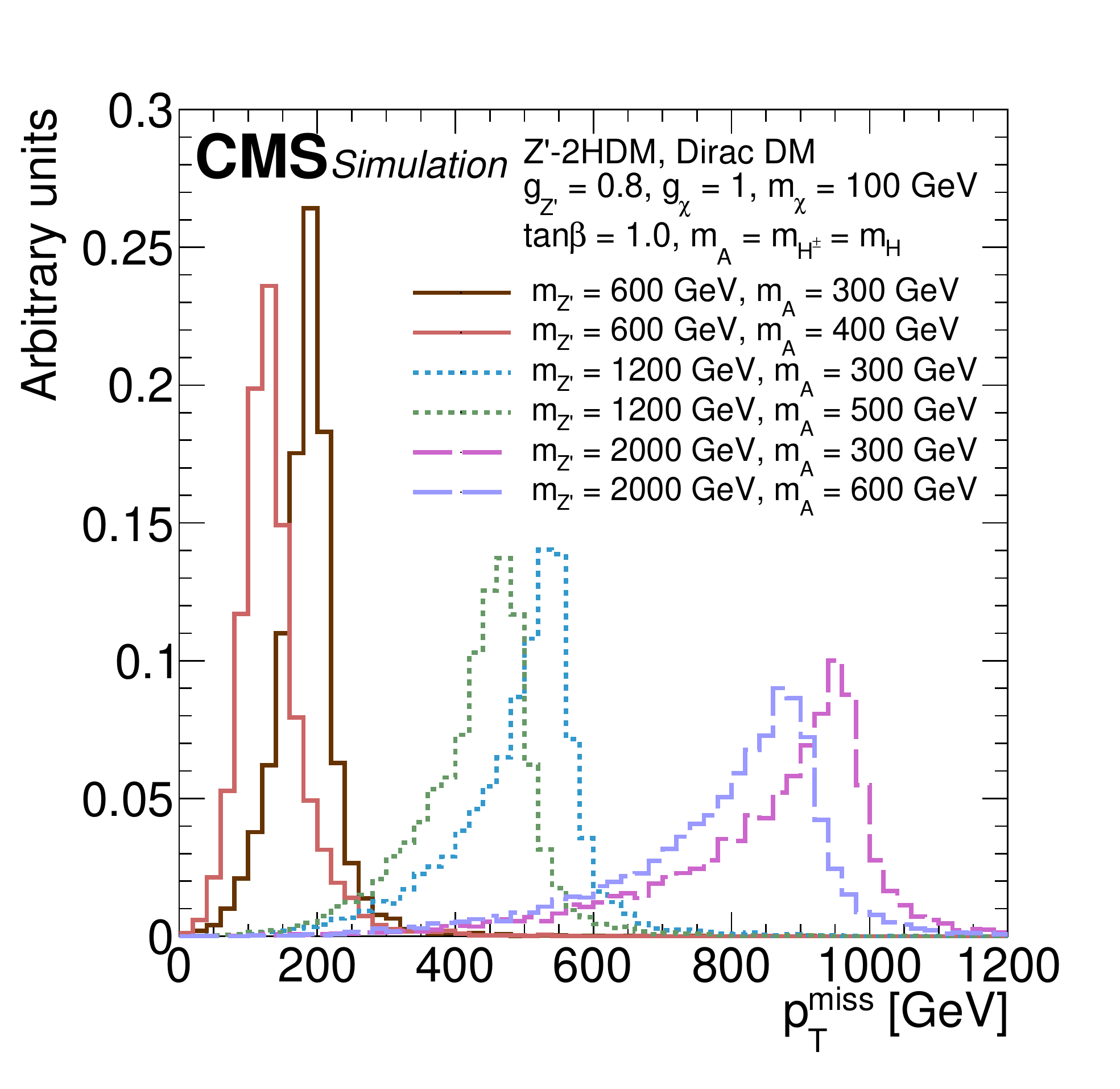}
    \includegraphics[width=0.475\textwidth]{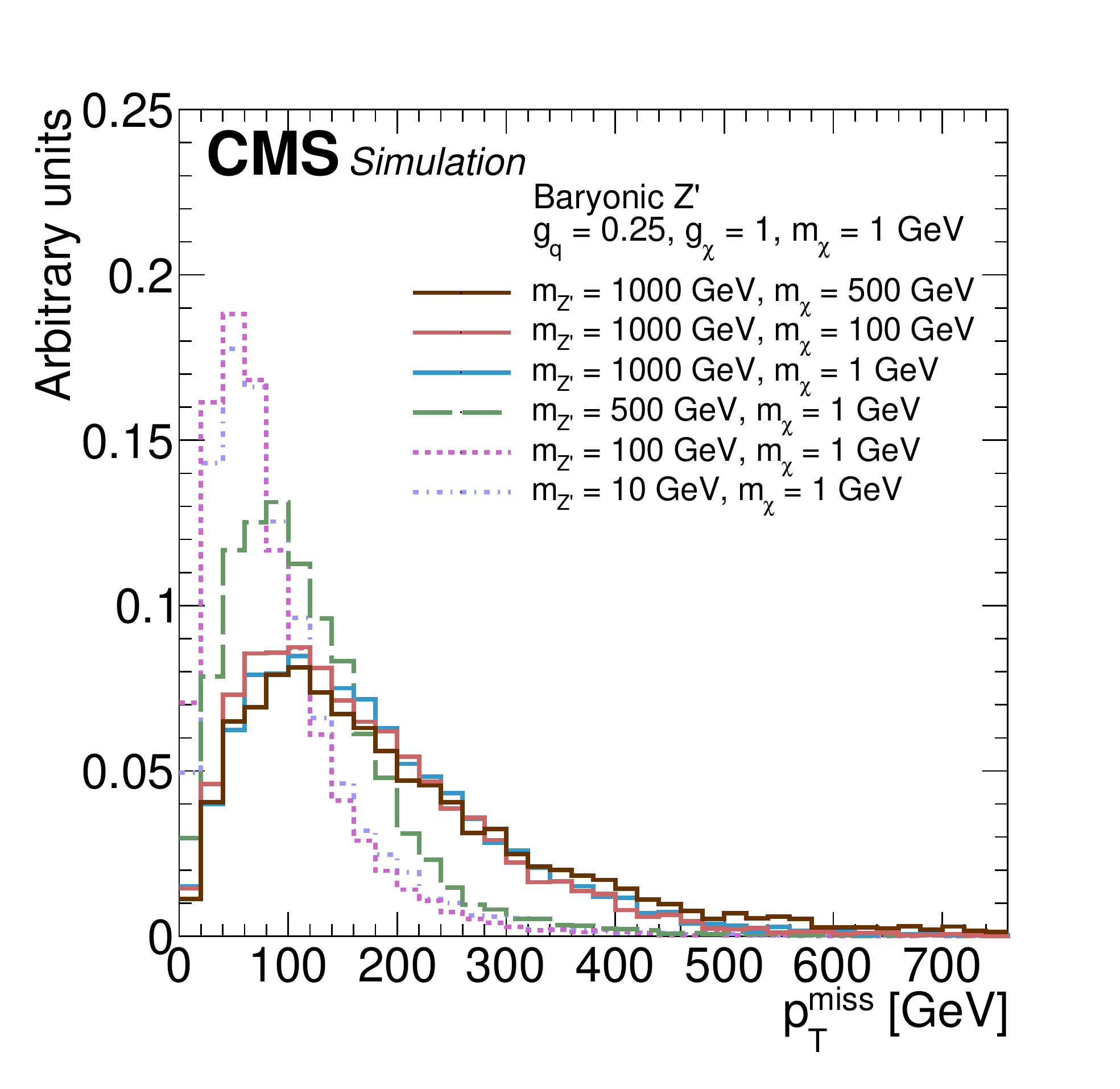}
  \caption{The distribution of \ptmiss at the generator level for the \PZp-2HDM (left), showing the dependence on the two main model parameters varied in the analysis, \mZp and \mA,
 and for the baryonic \PZp model (right), showing the variation of \ptmiss as a function of \mZp and \mchi. All other parameters of the models are fixed to the values specified in the text. The distributions are normalized to unit area.}
  \label{fig:genmet}
\end{figure}

For the baryonic \PZp model, \mZp between 100 and 2500\GeV and \mchi between 1 and 700\GeV are used for this study.
The \PZp-DM coupling is fixed to $g_{\PGc} = 1$ and the \PZp-quark coupling  is fixed to $\gq = 0.25$. The mixing angle between the baryonic Higgs boson and the SM-like Higgs boson is set to $\sin\theta = 0.3$, and the coupling between
the \PZp boson and \Ph is assumed to be proportional to \mZp. The branching fractions of the Higgs boson decays are altered for $\mZp \lesssim \mH/2$, because the decay $\Ph \to {\PZp\PZp}^{(*)}$ becomes kinematically accessible. Therefore the region $\mZp < 100\GeV$, for which the modification of the $\Ph$ branching fractions is sizable, is not considered in the analysis.
For both benchmark models, \Ph is assumed to have a mass of 125\GeV. A considerable amount of \ptmiss is expected, as shown in Fig.~\ref{fig:genmet}. 
The reason that the  \ptmiss spectrum is harder for the \PZp-2HDM is that the DM particles are produced via a resonant mechanism in this case, whereas for the baryonic  \PZp model they are not. The difference in shape becomes more marked as \mZp increases. In Fig.~\ref{fig:genmet} (right) it can be seen that the shape of the \ptmiss distribution is almost independent of \mchi in the baryonic \PZp model, and depends most strongly on \mZp.

Although the signal sensitivity in the \hbb channel is higher than in the other final states considered ($\PGg\PGg$, $\Pgt\Pgt$, $\PW\PW$, and  $\PZ\PZ$) because of the channel's large branching
fraction and
manageable background in the large-\ptmiss region, the statistical combination of all five decay modes is performed to improve the overall sensitivity. The \hgg and \hzz channels exhibit better resolution in the reconstructed Higgs boson invariant mass, while the \htt, \hww, and \hzz channels benefit from lower SM backgrounds, which results in a higher sensitivity for signals with a soft \ptmiss spectrum.

In the \hbb channel analysis, the \Ph is reconstructed from two overlapping \PQb jets. Thus different approaches are used for the two models, because of the difference in the average Lorentz boost of the Higgs boson, which is higher in the \PZp-2HDM than in the baryonic \PZp model. The Higgs boson is reconstructed using a jet clustering algorithm with a distance parameter of 0.8 for the \PZp-2HDM and 1.5 for the baryonic \PZp model. For the baryonic \PZp model, a simultaneous fit  of the distribution of the recoil variable in the signal region (SR) and the control regions (CRs) is performed to extract the signal. For the \PZp-2HDM, a parametric fit of the \PZp boson transverse mass is used to estimate the major backgrounds and to extract the signal.

The search in the \hgg channel~\cite{exo:monohggtt2016} uses a fit to the diphoton invariant mass distribution to extract the signal. This analysis is performed in two categories distinguished by the \ptmiss value, high ( $>$130\GeV) and low (50--130\GeV), in order to be sensitive to a large variety of possible signals.

The search in the \htt channel~\cite{exo:monohggtt2016} is based on the combination of the events for the three \Pgt\ lepton decay modes with the highest branching fractions: $\tauh\tauh$, $\Pgm\tauh$, and $\Pe\tauh$, where \tauh denotes a hadronically decaying \Pgt\ lepton. After requiring a \ptmiss ($>$105\GeV) in order to suppress the background sufficiently, the signal is extracted by performing a simultaneous fit
in the SR and in the CRs to the transverse mass of the Higgs boson reconstructed from the two $\Pgt$ leptons. In the \hww channel search, the fully leptonic decays of the two \PW\ bosons are considered, requiring one lepton to be an electron and the other to be a muon, in order to reduce the contamination from the $\PZ\to\EE$ and $\PZ \to \MM$ backgrounds.
The \hzz search is performed in the fully leptonic decay channel of the \PZ boson pair: $\Ph \to \PZ\PZ \to 4\ell$. The analysis strategy follows closely the measurement of the Higgs boson properties in the same channel~\cite{Sirunyan:2017exp}.

The paper is organized as follows. After a brief introduction of the CMS detector in Section~\ref{sec:detector}, the data and simulated event samples are described in Section~\ref{sec:sigbkgsim}. The event reconstruction and the analysis strategy for each Higgs boson decay mode used in the statistical combination are detailed in Sections~\ref{sec:recoandid} and~\ref{sec:eventselection}, respectively. The combination procedure and the main systematic uncertainties are described in Sections~\ref{combo} and~\ref{sec:systematics}, respectively. The results are presented in Section~\ref{sec:results}, and the paper is summarized in Section~\ref{sec:summary}.

\section{The CMS detector}\label{sec:detector}

The central feature of the CMS apparatus is a superconducting solenoid of 6\unit{m} internal diameter, providing a magnetic field of 3.8\unit{T}. Within the solenoid volume are a silicon pixel and strip tracker, a lead tungstate crystal electromagnetic calorimeter (ECAL), and a brass and scintillator hadron calorimeter (HCAL), each composed of a barrel and two endcap sections. Forward calorimeters, made of steel and quartz fibres, extend the pseudorapidity ($\eta$) coverage provided by the barrel and endcap detectors. Muons are detected in gas-ionization chambers embedded in the steel flux-return yoke outside the solenoid.

Events of interest are selected using a two-tiered trigger system~\cite{Khachatryan:2016bia}. The first level, composed of custom hardware processors, uses information from the calorimeters and muon detectors to select events at a rate of around 100\unit{kHz} in a time of less than 4\mus. The second level, known as the high-level trigger, consists of a farm of processors running a version of the full event reconstruction software optimized for fast processing, and reduces the event rate to around 1\unit{kHz} before data storage.

 A more detailed description of the CMS detector, together with a definition of the coordinate system used and the relevant kinematic variables, can be found in Ref.~\cite{Chatrchyan:2008zzk}.

The $\Pp\Pp$ collision data were collected at $\sqrt{s}$ = 13\TeV in 2016. The time spacing between adjacent bunches of 25\unit{ns} leads to an average number of  $\Pp\Pp$ interactions per bunch crossing of 23 assuming the $\Pp\Pp$ inelastic cross section of 69.2\unit{mb}~\cite{Sirunyan:2018nqx}. The integrated luminosity of the  data sample used in all the analyses described in this paper corresponds to 35.9\fbinv, after imposing data quality requirements.

\section{Signal and background simulation} \label{sec:sigbkgsim}

Signal samples for the five Higgs boson decay modes are generated at leading order (LO) in perturbative quantum chromodynamics (QCD) using the \MGvATNLO~v2.3.0
generator~\cite{Alwall:2011uj,Alwall:2014hca}, for both the \PZp-2HDM and baryonic \PZp model~\cite{Abercrombie:2015wmb}. The Higgs boson is treated as a stable particle during the generation, and its decays are described subsequently using \PYTHIA~8.212~\cite{pythia8}.

A detailed description of the simulated samples used for the \hbb, \hgg, and \htt analyses can be found in Refs.~\cite{exo:monohbb2016,b2g:monohbb2016,exo:monohggtt2016}. The production of a Higgs boson in association with a \PZ boson decaying to a pair of neutrinos is an irreducible background for all the final states considered.
Other Higgs boson backgrounds originating from gluon-gluon fusion ($\Pg\Pg$F) and vector boson fusion (VBF) production modes are small. These backgrounds are simulated at next-to-LO (NLO) in QCD with \POWHEG{} v2~\cite{Nason:2004rx, Frixione:2007vw, Alioli:2010xd}.

The main nonresonant backgrounds in the \hww analysis are from the continuum $\PW\PW$, single top quark, and top quark pair production.
The continuum $\PW\PW$ production is simulated in different ways: \POWHEG~\cite{Melia:2011tj}
is used to generate $\qqbar \to \PW\PW$ events at NLO precision, whereas $\Pg\Pg \to \PW\PW$ events are  generated at LO using \MCFM v7.0~\cite{Melnikov:2006kv,Campbell:2011bn,Campbell:2015qma}. The simulated $\qqbar \to \PW\PW$ events are reweighted to reproduce the \ptww distribution from the \pt-resummed calculation at next-to-NLO (NNLO) plus next-to-next-to-leading logarithmic precision~\cite{Meade:2014fc,Jaiswal:2014yba}. The LO $\Pg\Pg \to \PW\PW$ cross section, obtained directly from \MCFM, is further corrected to NNLO precision via a $K$ factor of 1.4~\cite{Caola:2015rqy}.
Single top quark, \ttbar, $\PW\PZ$, and $\PW\PGg^*$ backgrounds are generated at NLO with \POWHEG. Drell--Yan (DY) production of $\PZ/\PGg^{*}$ is generated at NLO using \MGvATNLO, and the \pt spectrum of the dilepton pairs is reweighted to match the distribution observed in dimuon events in data.  Other multiboson processes, such as $\PW\PGg$, $\PZ\PZ$, and $\PV\PV\PV$ ($\PV = \PW$ or \PZ), are generated at NLO with \MGvATNLO. All samples are normalized to the latest available theoretical cross sections, NLO or higher ~\cite{Campbell:2011bn,Melnikov:2006kv,Khachatryan:2016mqs}.

In the \hzz analysis, the SM production mechanism constitutes a major background because this has the same experimental signature and satisfies the low \ptmiss threshold used in the analysis. It is simulated with \POWHEG~\cite{Alioli:2008gx,Nason:2004rx,Frixione:2007vw} in four main production modes: $\Pg\Pg$F, including quark mass effects~\cite{Bagnaschi:2011tu}; VBF~\cite{Nason:2009ai};
  associated production with a top quark pair ($\ttH$)~\cite{Hartanto:2015uka}; and associated production with a vector boson ($\PW\Ph$, $\PZ\Ph$), using the \textsc{minlo hvj}~\cite{Luisoni:2013kna} extension of \POWHEG.
In all cases, the Higgs boson is forced to decay via the \HZZfl ($\ell = \Pe$, $\PGm$, or $\PGt$) channel. The description of the decay of the Higgs boson to four leptons is obtained using the \textsc{JHUgen}~7.0.2 generator~\cite{Gao:2010qx,Bolognesi:2012mm}. In the case of $\PZ\Ph$ and $\ttH$ production, the Higgs boson is allowed to decay as $\Ph \to \cPZ \cPZ \to 2\ell+\PX$, such that four-lepton events where two leptons originate from the decay of the associated $\PZ$ boson or top quarks are also taken into account in the simulation. The cross sections for the processes involving SM Higgs boson production
are taken from Ref.~\cite{yr4}.

All processes are generated using the NNPDF3.0~\cite{Ball:2014uwa} parton distribution functions (PDFs), with the precision matching the parton-level generator precision. The \PYTHIA generator with the underlying event tune CUETP8M1~\cite{Khachatryan:2015pea} is used to describe parton showering and fragmentation. The detector response is simulated using a detailed
description of the CMS apparatus, based on the \GEANTfour package~\cite{Agostinelli:2002hh}.
Additional simulated $\Pp\Pp$ minimum bias interactions in the same or adjacent bunch crossings (pileup) are added to the hard scattering event, with the multiplicity distribution adjusted to match that  observed in data.

\section{Event reconstruction} \label{sec:recoandid}

The particle-flow (PF) algorithm~\cite{Sirunyan:2017ulk} aims to reconstruct and identify each individual particle in an event, with an optimized combination of information from the various elements of the CMS detector. The energy of photons is obtained from the ECAL measurement. The energy of electrons is determined from a combination of the electron momentum at the primary interaction vertex as determined by the tracker, the energy of the corresponding ECAL cluster, and the energy sum of all bremsstrahlung photons spatially compatible with originating from the electron track~\cite{Khachatryan:2015hwa}. The energy of muons is obtained from the curvature of the corresponding track. The energy of charged hadrons is determined from a combination of their momentum measured in the tracker and the matching ECAL and HCAL energy deposits, corrected for zero-suppression effects and for the response function of the calorimeters to hadronic showers. Finally, the energy of neutral hadrons is obtained from the corresponding corrected ECAL and HCAL energies.

Electron candidates are required to have
$\abs{\eta}<2.5$. Additional requirements are applied to reject electrons originating from photon
conversions in the tracker material or jets misreconstructed as electrons.
Electron identification criteria rely on observables sensitive to the bremsstrahlung along the
electron trajectory and on the geometrical and momentum-energy matching between the electron track
and the associated energy cluster in the ECAL, as well as on the ECAL shower
shape observables and association with the primary vertex.

Muon candidates are reconstructed within $\abs{\eta}<2.4$
by combining information from the silicon tracker and the muon system.
Identification criteria based on the number of measurements in the
tracker and in the muon system, the fit quality of the muon track, and its consistency with its
origin from the primary vertex are imposed on the muon candidates to reduce the misidentification
rate.

For each event, hadronic jets are clustered from PF candidates using the infrared- and collinear-safe anti-\kt algorithm~\cite{Cacciari:fastjet2,Cacciari:fastjet1}, with a distance parameter of 0.4 (AK4 jets) or 0.8 (AK8 jets). Jet momentum is determined as the vectorial sum of all particle momenta in the jet, and is found from simulation to be, on average, within 5 to 10\% of the true momentum over the entire \pt spectrum and detector acceptance. Pileup interactions can result in additional spurious contributions to the jet momentum measurement from tracks and calorimetric energy depositions. To mitigate this effect, tracks identified to be originating from pileup vertices are discarded and a correction based on the jet area~\cite{Cacciari:2007fd} is applied to account for the neutral pileup particle contributions. Jet energy corrections are derived from simulation to bring the measured response of jets to that of particle-level jets on average. In situ measurements of the momentum balance in dijet, photon+jet,
 Z+jet, and multijet events are used to account for any residual differences in the jet energy scale (JES) between data and simulation~\cite{Khachatryan:2016kdb}. The jet energy resolution (JER) amounts typically to 15\% at $\pt = 10\GeV$, 8\% at 100\GeV, and 4\% at 1\TeV. Additional selection criteria are applied to remove jets potentially dominated by anomalous contributions from various subdetector components or reconstruction failures~\cite{CMS-PAS-JME-10-003}.

At large Lorentz boosts, the two \PQb\ quarks from the Higgs boson decay
may produce jets that overlap and make their individual reconstruction difficult.
In this case, either the AK8 jets or larger-area jets clustered from PF candidates using the
Cambridge--Aachen algorithm~\cite{Wobisch:1998wt,CMS-PAS-JME-09-001} with a distance parameter of
1.5 (CA15 jets) are used. To reduce the impact of particles arising from pileup interactions when reconstructing AK8 or CA15 jets,
the four-vector of each PF candidate matched to the jet is scaled with a weight calculated
with the pileup-per-particle identification algorithm~\cite{Bertolini:2014bba} prior to the clustering.
The CA15 jets are also required to be central ($\abs{\eta} <2.4$).
The ``soft-drop'' jet grooming algorithm~\cite{Larkoski:2014wba} is applied to remove soft,
large-angle radiation from the jets. The mass of a groomed AK8 or CA15 jet are referred to as $m_\text{SD}$.

To identify jets originating from \PQb quark fragmentation (\PQb jets), two \PQb tagging algorithms are used.
The combined secondary vertex (CSVv2)~\cite{Sirunyan:2017ezt} and the combined multivariate analysis (cMVAv2) algorithms~\cite{Sirunyan:2017ezt} are used to identify AK4 jets originating from \PQb quarks by their characteristic displaced vertices. For the AK8 jets, subjets inside the jet are required to be tagged as \PQb jets using the CSVv2 algorithm.
A likelihood for the CA15 jet to contain two
\PQb quarks is derived by combining the information from the primary and secondary vertices
and tracks in a multivariate discriminant optimized to distinguish CA15 jets originating
from the \hbb  decay from those produced by
energetic light-flavor quarks or gluons~\cite{exo:monohbb2016}.

Hadronically decaying $\Pgt$ leptons are reconstructed from jets using the
hadrons-plus-strips algorithm~\cite{hpssource}.
This algorithm uses combinations of reconstructed charged hadrons and energy deposits in the ECAL to identify the three most common hadronic $\Pgt$ lepton decay modes: 1-prong, 1-prong+$\PGpz$(s), and 3-prong.
The \tauh candidates are further required to satisfy the isolation criteria with an efficiency of 65 (50)\% and a misidentification probability of 0.8 (0.2)\% in the \tauhtauh\ (\tauetauh\ or \taumutauh) channel.

The \ptvecmiss is reconstructed as the negative vectorial sum of all PF particle candidate momenta projected on the plane transverse to the beams. Since the presence of pileup induces a degradation of the \ptmiss measurement (\ptmiss resolution varies almost linearly from 15 to 30\% as the number of vertices increases from 5 to 30 ~\cite{JME-17-001}), affecting mostly backgrounds with no genuine \ptmiss,  an alternative definition of \ptmiss that is constructed only using the charged PF candidates (``tracker \ptmiss'') is used in the \hww analysis. In the rest of the paper, \ptmiss corresponds to the PF \ptmiss, unless specified otherwise.

\section{Analysis strategy} \label{sec:eventselection}

In this section we briefly discuss the analysis strategies in the previously published~\cite{b2g:monohbb2016,exo:monohbb2016,exo:monohggtt2016} \hbb, \hgg, and \htt, channels, and provide full descriptions of the new analyses in the \hww and \hzz decay channels. The summary of all the decay channels contributing to the
combination is presented in Table~\ref{tab:decaymodes}.

\begin{table*}[t]
    \centering
    \topcaption{\label{tab:decaymodes}Summary of the individual channels entering the combination. Analyses are categorized based on the model, \ptmiss selection, and subsequent decay products listed here. The categorization is the same for both the \PZp-2HDM and the Baryonic \PZp model for all decay channels except, as indicated, \hbb. A dash (``\NA") in the last column implies that the analysis is presented in this paper.}
     \begin{tabular}{  l  c  c  c  }
        \hline
             Decay channel & Final state or category &  Reference  \\ \hline
               \multirow{2}{*}{\hbb} & AK8 jet (\PZp-2HDM) & \protect\cite{b2g:monohbb2016} \\
                                     & CA15 jet (Baryonic \PZp) & \protect\cite{exo:monohbb2016}\\[\cmsTabSkip]
               \multirow{2}{*}{\hgg} & $\ptmiss \in$ 50--130\GeV  & \protect\cite{exo:monohggtt2016} \\
                                     & $\ptmiss >130\GeV$  & \protect\cite{exo:monohggtt2016} \\[\cmsTabSkip]
               \multirow{3}{*} {\htt}
                                     & \tauhtauh    &\protect\cite{exo:monohggtt2016}\\
                                     & \taumutauh   &\protect\cite{exo:monohggtt2016}\\
		                     & \tauetauh   &\protect\cite{exo:monohggtt2016}\\[\cmsTabSkip]
               \hww                  & $\Pe\Pgn \Pgm \Pgn$      &  \NA \\[\cmsTabSkip]
               \multirow{3}{*}{\hzz} & $4\Pe$  &  \NA \\
                                     & \mmmm   &  \NA \\
                                     & \eemm   &  \NA \\

          \hline
         \end{tabular}
\end{table*}

\subsection{The \texorpdfstring{$\hbbp + \ptmiss$}{h(bb) + pTmiss} channel}

The events used in this final state are selected using a triggers that require large amount ($>90$ or $>120\GeV$) of \ptmiss, or \mht  defined as the magnitude of the vectorial sum of the transverse momenta of all jets with $\pt > 20\GeV$ in an event. The trigger selection is 96 (100)\% efficient for events that subsequently have $\ptmiss > 200$ (350)\GeV in the off-line reconstruction.  As can be seen in  Fig.~\ref{fig:genmet}, the Lorentz boosts of the Higgs boson are different for the \PZp-2HDM and baryonic \PZp model. The events with large boost in the \PZp-2HDM are reconstructed using a large-radius AK8 jet with $\pt > 200\GeV$ and $\abs{\eta} < 2.4$. In addition, the \hbb topology is selected by requiring at least one subjet of the AK8 jet to be  \PQb tagged.
The analysis considers separately two categories, distinguished by the number of \PQb tagged subjets in the event, one or two, the latter being the high-purity category with higher sensitivity. For events with lower boost in the baryonic \PZp model, Higgs boson candidates are reconstructed using CA15 jets.

To select the \hbb candidates using the AK8 jet, one or both subjets are required to pass the loose \PQb tagging criteria, which has an efficiency of 85\%, and a misidentification rate of about 10\%
for jets originating from light-flavor quarks or gluons. In the case of the CA15 jets, a multivariate double \PQb tagging algorithm~\cite{Sirunyan:2017ezt} is used to discriminate the signal
from the background of light-flavor jets~\cite{exo:monohbb2016}, with an efficiency of 50\% and a misidentification rate of 10\%. The AK8 (CA15) analysis requires the Higgs boson candidate mass to be in the 105--135 (100--150)\GeV range to reduce nonresonant backgrounds.
The difference in the two mass window requirements is primarily driven by the differences in the performance of the two algorithms and in the jet mass resolutions. For both analyses, the mass window was chosen to maximize the signal sensitivity. In order to further reduce the background contributions from \wj and \ttbar production, events with an electron, muon, photon ($\pt > 10\GeV$), or \tauh ($\pt >18\GeV$) candidates passing loose identification and isolation criteria are vetoed. Furthermore, in the AK8 analysis, the number of additional \PQb tagged AK4 jets with $\pt > 20\GeV$ is required to be
 zero, while in the CA15 analysis, the number of AK4 jets with $\pt >30\GeV$, well-separated from the CA15 jet in the event, is required to be at most one. The sensitivity of the analyses is further enhanced by using jet substructure variables. The full details of the event selection for the AK8 and CA15 jet analyses can be found in Refs.~\cite{b2g:monohbb2016} and \cite{exo:monohbb2016}, respectively.

\subsection{The \texorpdfstring{$\hggp+\ptmiss$}{h(gamma gamma)+pTmiss} channel}

Signal candidate events in the \hgg analysis are selected using a diphoton trigger with asymmetric \pt thresholds of 30 and 18\GeV on the leading and subleading photons, respectively, and loose identification and isolation requirements imposed on both photon candidates. The diphoton invariant mass is further required to exceed 90\GeV.

Slightly higher thresholds of 30 (20)\GeV on the leading (subleading) photon \pt and of 95\GeV on the diphoton mass are used offline. The photon candidates are required to pass the isolation criteria if the spatial distance in $\eta$-$\phi$ plane ($\Delta R = \sqrt{\smash[b]{(\Delta\eta)^2+(\Delta\phi)^2}} $) between the two photons exceeds 0.3. The isolation selection is not used for photons that are coming from the decay of a highly Lorentz-boosted Higgs boson, as the two photons are likely to be found in the isolation cone of one another. The analysis is performed in two categories distinguished by the value of \ptmiss: high-\ptmiss ($>$130\GeV) and low-\ptmiss (50--130\GeV).

The multijet background, with a large \ptmiss in an event originating from the mismeasurement of the energy of one or more jets, is reduced by allowing at most two jets with $\pt > 30\GeV$.
To suppress the contribution from the multijet background, the azimuthal separation between the direction of any jet with $\pt > 50\GeV$ and \ptvecmiss is required to exceed 0.5 radians.
Finally, to select signal-like events with the DM particles recoiling against the Higgs boson, the azimuthal separation between  \ptvecmiss  and the direction of the Higgs boson candidate reconstructed from the diphoton system is required to exceed 2.1 radians.
More details of the event selection can be found in Ref.~\cite{exo:monohggtt2016}.

\subsection{The \texorpdfstring{$\http+\ptmiss$}{h(tau tau) + pTmiss} channel}

In the \htt analysis, the three final states with the highest branching fractions are analyzed: \tauhtauh, \taumutauh, and \tauetauh. The events are selected online with a  trigger requiring the presence of two isolated  \tauh candidates in the $\tauh\tauh$ final state, and a single-muon (single-electron) trigger in the \taumutauh\ ($\tauetauh$) final state. Electron, muon, and \tauh candidates passing the identification and isolation criteria are combined to reconstruct a Higgs boson candidate in these three final states. The signal events are then selected with the requirements: $\ptmiss > 105\GeV$
and visible \pt of the \tautau system $> 65\GeV$. To ensure that the \tautau system originates from the Higgs boson, the visible mass of the \tautau system is required to be less than 125\GeV. In order to reduce the contribution from multilepton and \ttbar backgrounds, the events are vetoed if an additional electron, muon, or a \PQb tagged jet is present. More details of the event selection can be found in Ref.~\cite{exo:monohggtt2016}.

\subsection{The \texorpdfstring{$\hwwp+\ptmiss$}{h(WW) + pTmiss} channel}\label{sec:WWanalysls}

The search in the \hww decay channel is performed in the fully leptonic, opposite-sign, different-flavor ($\Pe\PGm$) final state, which has relatively low backgrounds. The presence of the neutrinos and the DM particles escaping detection results in large \ptmiss in signal events. The selected $\Pe\Pgm+\ptmiss$ events include a contribution from the $\hww \to \Pgt\Pgt\nu_{\Pgt}\nu_{\Pgt}$ process with both \Pgt leptons decaying leptonically. Several background processes can lead to the same final state, dominated by \ttbar and $\WW$ production.

Online, events are selected using a suite of single- and double-lepton triggers. In the offline selection, the leading (subleading) lepton  is  required to have $\pt > 25$ (20)\GeV. Electron and muon candidates are required to be well-identified and isolated to reject the background from leptons inside jets. Backgrounds from low-mass resonances are reduced by requiring the dilepton invariant mass (\mll) to exceed 12\GeV, while backgrounds with three leptons in the final state are reduced by vetoing events with an additional  well-identified lepton with $\pt >10 \GeV$.  The \ptmiss in the event is required to exceed 20\GeV in order to reduce the contribution from instrumental backgrounds and \dytt\ decays. To suppress the latter background, the \pt of the dilepton system
is required to be greater than 30\GeV and the transverse mass of the dilepton and \ptvecmiss system, \mTH, is required to be greater than 40\GeV.
In order to reduce the $\PZ/\PGg^* \to \EE, \MM$ or $\Pgt^{+}\Pgt^{-}$ background with \ptmiss originating either from $\PGt$ lepton decays or from mismeasurement
of the energies of \Pe, $\mu$ or additional jets, a variable \pTmissproj~\cite{Sirunyan:2018egh} is introduced. This is defined as the projection of  ${\ptvecmiss}$ in the plane
transverse to the direction of the nearest lepton, unless this lepton is situated in the opposite hemisphere to $ {\ptvecmiss}$, in which case \pTmissproj is taken to be \ptmiss itself.
 A selection using this variable efficiently rejects  $\PZ/\PGg^* \to \ell\ell$ background events, in which the $ {\ptvecmiss}$ is preferentially aligned with leptons.
 Since the \ptmiss resolution is degraded by pileup, a quantity \pTmissmp is defined as the smaller of the two \pTmissproj values: the one based on all the PF candidates in the event, and the one based only on the reconstructed tracks originating from the primary vertex. A requirement $\pTmissmp > 20\GeV$ is effective in suppressing the targeted background. The above requirements define the event preselection.

\begin{figure}[htbp]
  \centering
  \includegraphics[width=0.475\textwidth]{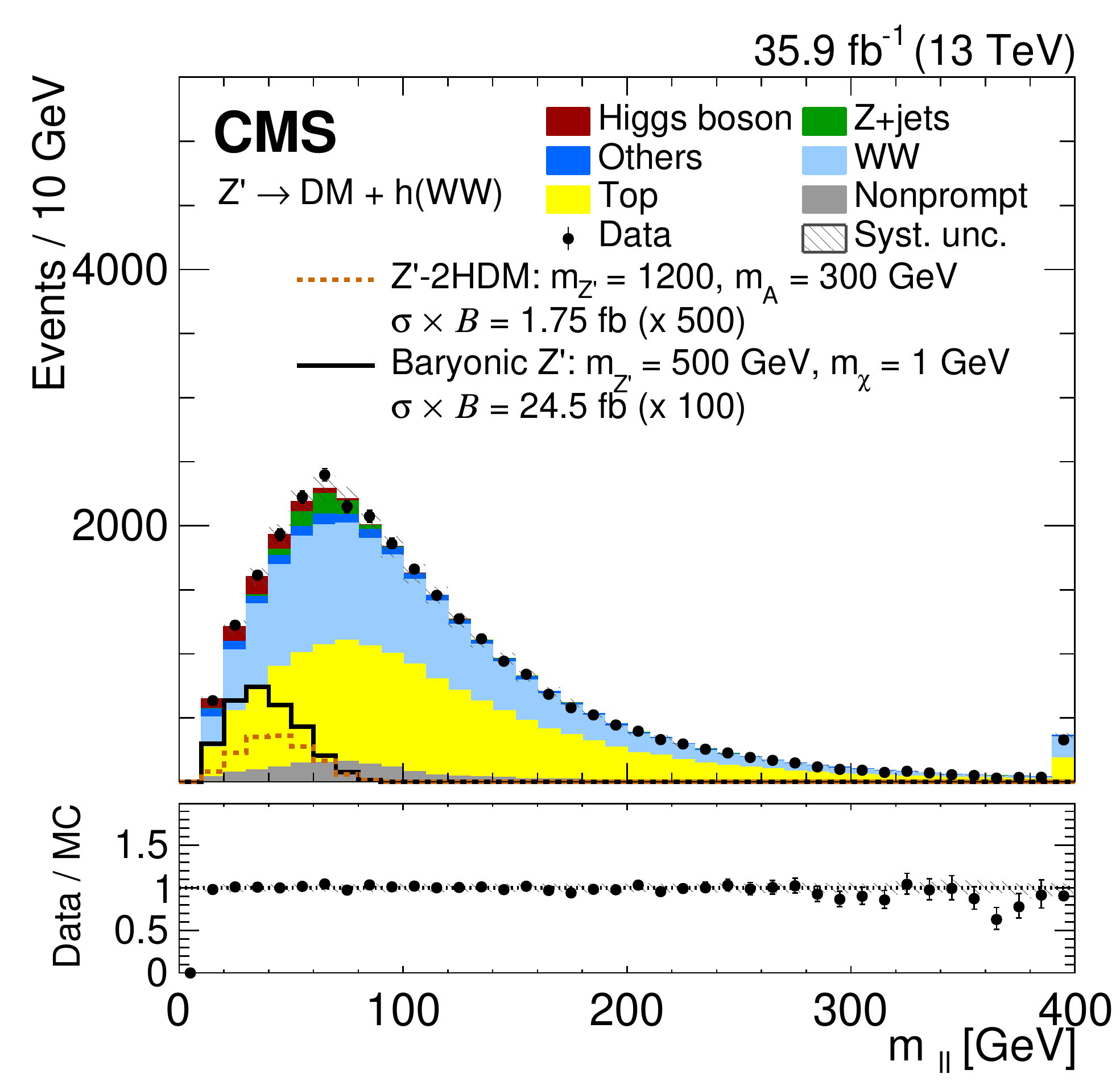}
   \includegraphics[width=0.475\textwidth]{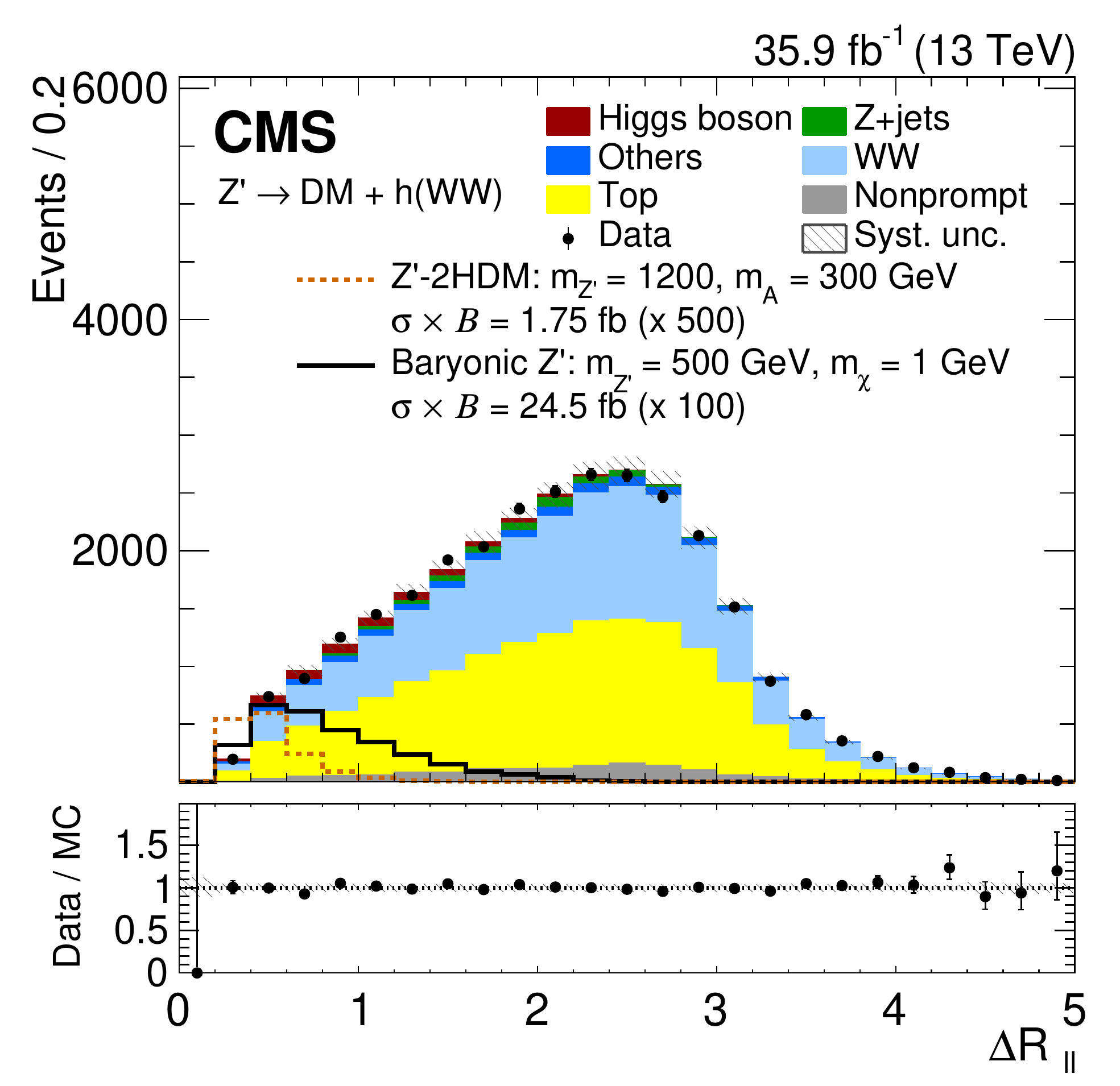}
     \caption{The distribution of \mll (left) and \delRll (right) after the preselection, expected from MC simulation (stacked histograms) and observed  in data (points with vertical bars).  The systematic uncertainties, discussed in Section~\protect\ref{subsec:systWW}, are shown by the hatched region. Two signal benchmarks, corresponding to the \PZp-2HDM (dotted orange line) and baryonic \PZp (solid black line) model are superimposed. The signal is normalized to the product of cross section and $\mathcal{B}$, where $\mathcal{B}$ represents the $\Ph \to \WW$ branching fraction. The signal distributions are scaled up by a factor 500 (100) for the \PZp-2HDM (baryonic \PZp model), to make them more visible. The lower panel shows the ratio of the data to the predicted SM background. }
  \label{fig:cuts}
\end{figure}

The expected signal significance is enhanced by introducing two additional selections: $\mll < 76\GeV$ and the distance in $\eta$-$\phi$ space between the two leptons $\delRll < 2.5$, as illustrated in Fig.~\ref{fig:cuts}.
The first requirement exploits the fact that the invariant mass of the leptons coming from the \hww decay tends to be low because of the presence of the two neutrinos in the decay chain and of the scalar nature of the Higgs boson. The second requirement utilizes the fact that the Higgs boson in signal events recoils against the DM particles and is highly boosted.

\subsubsection{Background estimation} \label{sec:WWExtraction}

Since full kinematic reconstruction of the Higgs boson mass and \pt is impossible in this decay channel because of the presence of undetected neutrinos and DM particles, to maximize the sensitivity of the search, a boosted decision tree (BDT) multivariate classifier has been trained for each of the two signal models. The BDT exploits the following input variables:
\begin{itemize}
\item transverse masses: \mTH, \mTwl, \mTwt;
\item lepton transverse momenta: \ptll, \ptla, \ptlb;
\item missing transverse momenta: PF \ptmiss, \tkmet, \pTmissmp;
\item angular variables: \delRll, \delphill, \delphilmetl, \delphilmett; and
\item dilepton invariant mass: \mll.
\end{itemize}
Here,  $\mTwll = \sqrt{\smash[b]{ 2 \pt^{\ell_{i}} \ptmiss(1 - \cos \delphilmetll)}}$, where $i=1$ ($i=2$) defines the transverse mass of \ptvecmiss and the leading (subleading) lepton in the event,  and \delphill\ is the azimuthal angle between the directions of the two lepton momenta.

For both benchmark models, the BDT training considers processes with two prompt leptons and genuine \ptmiss (\WW, \ttbar, \tw, and \hww production) as the backgrounds. For the \PZp-2HDM (baryonic \PZp) model, simulated signal samples with $\mA=300\GeV$ ($\mChi=1\GeV$) with various values of \mZp have been used for training. The chosen signal points correspond to the region of maximum sensitivity of the \hww analysis for both models.

The main background processes arise from top quark (\ttbar and single top quark production, mainly \tw), nonresonant \WW events, and nonprompt leptons.
The contribution of nonprompt-lepton background in the SR is determined entirely from data, while the contributions of the top quark, \WW, and \dytt background are estimated using simulated samples.
 The normalizations of simulated backgrounds are obtained using dedicated CRs that are included in the maximum-likelihood fit used to extract the signal, together with the SR. Smaller backgrounds, \wz and \wgs,  are estimated using simulation after applying a normalization factor estimated in the respective CRs. The \wz CR is defined by requiring the presence of two opposite-sign, same-flavor leptons, compatible with the decay of a \PZ boson and one additional lepton of a different flavor, consistent with originating from a \PW\ boson decay. In the \wgs CR, the two leptons produced by the decay of the virtual photon are required to have $\pt > 8\GeV$ and be isolated.
Since the two leptons may be close to each other, the isolation is computed without taking into account the contribution of lepton tracks falling in the isolation cone. An additional lepton consistent with originating from the \PW\ decay is required. The \wz and \wgs CRs are not used in the maximum-likelihood fit; instead, the normalization scale factors are extracted and directly applied to the corresponding simulated samples. The remaining backgrounds from diboson and triboson production are estimated directly from simulation.

The $\ggww$ and $\qqww$ backgrounds are estimated from simulation normalized as discussed in Section~\ref{sec:sigbkgsim}.
The main feature of these processes is that, as the two \PW\ bosons do not originate in a decay of the Higgs boson, their invariant mass does not peak at the Higgs boson mass. For this reason, events in the corresponding CR are required to have a large dilepton invariant mass, achieved by inverting the SR $\mll < 76\GeV$ requirement.

The estimation of the top quark background is performed in two steps. First, a top quark enriched CR is defined to measure a scale factor quantifying the difference in the \PQb tagging efficiencies and mistag rates in data and simulation. This CR is obtained
from the SR selection by inverting the \PQb tagged jet veto. In second step, the scale factor  is applied to the corresponding simulated samples with a weight per event that depends on the number, flavor, and kinematic distributions of jets.

The \wj production contributes as a background in the \hww analysis when a jet is misidentified as a lepton. A CR is defined to contain events with one isolated lepton and another lepton candidate that fails the nominal isolation criteria, but passes a looser selection. The probability for a jet satisfying this looser selection to pass the nominal one is estimated from data in an independent sample dominated by nonprompt leptons from multijet production.  This probability is parameterized as a function of the \pt and $\eta$ of the lepton and applied to the events in the CR. In order to estimate the nonprompt lepton contamination in the SR, a validation region enriched in nonprompt leptons is defined with the same requirement as the SR, but requiring same-sign \emu pairs. The maximum discrepancy between data and prediction in the validation region, amounting to $\approx$$30\%$, is taken as the uncertainty in the \wj background prediction.

The \dytt background is estimated from simulation, after reweighting the \PZ boson \pt spectrum to match the distribution measured in data. The normalization of the simulated sample is estimated from data using events in the $\mTH < 40\GeV$ region. A normalization factor is then extracted from this region and applied to the SR.

The main difference between the present analysis and the measurement of the SM Higgs boson properties in the same channel~\cite{Sirunyan:2018egh} is in the signal extraction method. The latter analysis uses a multidimensional fit to the \mTH, \mll, and \ptlb~distributions, whereas a fit to the BDT discriminant distribution is used in the present analysis.

\subsection{The \texorpdfstring{$\hzzp+\ptmiss$}{h(ZZ) + pTmiss} channel}

The search in the \hzz channel is performed in all-leptonic final states. Each of the \PZ bosons decays to a pair of leptons (electrons or muons, including those coming from leptonic \PGt decays) resulting in a four-lepton signature. The main advantages of the \hzzfl over other Higgs boson decay modes are that the Higgs boson candidates can be fully reconstructed, with an excellent mass resolution, and the backgrounds are easily controlled.
On the other hand, this channel suffers from a relatively small branching fraction compared to most of other Higgs boson decay channels. The three different final states (four electrons, four muons, and two electrons and two muons) are analyzed individually and then combined to obtain final results. The selection of the \HZZfl events follows closely that used in the measurement of the Higgs boson properties in the four-lepton channel, based on the same data set~\cite{Sirunyan:2017exp}.

The signal event topology is defined by the presence of four charged leptons ($4\Pe$, $4\Pgm$, or $2\Pe2\Pgm$) and significant \ptmiss produced by the undetected DM particles. The events are selected online with triggers requiring the presence of two isolated leptons ($\Pe\Pe$, $\PGm\PGm$, or $\Pe\PGm$), with asymmetric \pt thresholds of 23 (17)\GeV on the leading and 12 (8)\GeV on the subleading electron (muon). Dilepton triggers account for most of the signal efficiency in all three final states. In order to maximize the signal acceptance,
trilepton triggers with lower \pt thresholds and no isolation requirements are added, as well as single-electron and single-muon triggers with isolated lepton  \pt thresholds of 27 and 22\GeV, respectively~\cite{Sirunyan:2017exp}.

The reconstruction and selection of the Higgs boson candidates proceeds first by selecting two \PZ boson candidates, defined as pairs of opposite-sign, same-flavor leptons ($\Pep\Pem$, $\Pgmp\Pgmm$) passing the selection criteria and satisfying
$12 < m_{\ell\ell(\PGg)} < 120\GeV$, where the \PZ boson candidate mass  $m_{\ell\ell(\PGg)}$ includes the contribution of photons identified as coming from final-state radiation~\cite{Sirunyan:2017exp}. The \ZZ candidates are then defined as pairs of \PZ boson candidates not sharing any of the leptons. The \PZ candidate with the reconstructed mass closest to the
nominal \PZ boson mass ~\cite{Tanabashi:2018oca} is denoted as ${\PZ_1}$, and the other one is denoted as ${\PZ_2}$.
All the leptons used to select the ${\PZ_1}$ and ${\PZ_2}$ candidates must be separated by $\Delta R(\ell_i, \ell_j) > 0.02$.

The leading (subleading) of the four leptons must have $\pt> 20$ (10)\GeV, and the ${\PZ_1}$ candidate must have a reconstructed mass $m_{\PZ_1}$ above 40\GeV. In the $4\Pe$ and $4\Pgm$ channels, if an alternative $\PZ_i\PZ_j$ candidate based on the same four leptons is found, the event is discarded if $m_{\PZ_i}$ is
closer to the nominal $\PZ$ boson mass than $m_{\PZ_1}$.
This requirement rejects events
with an on-shell $\PZ$ boson produced in association with a low-mass dilepton resonance. In order to suppress the contribution of QCD production of low-mass dilepton resonances, all four opposite-sign pairs that can be built with the four leptons
(regardless of the lepton flavor) must satisfy $m_{\ell_i\ell_j} > 4\GeV$ and the four-lepton invariant mass must satisfy $\mllll > 70\GeV$. If more than one \ZZ candidate passes the selection, the one with the highest value of the scalar \pt sum of four leptons is chosen. The above requirements define the event preselection.

\begin{figure}[!htb]
\begin{center}
\includegraphics[width=0.475\linewidth]{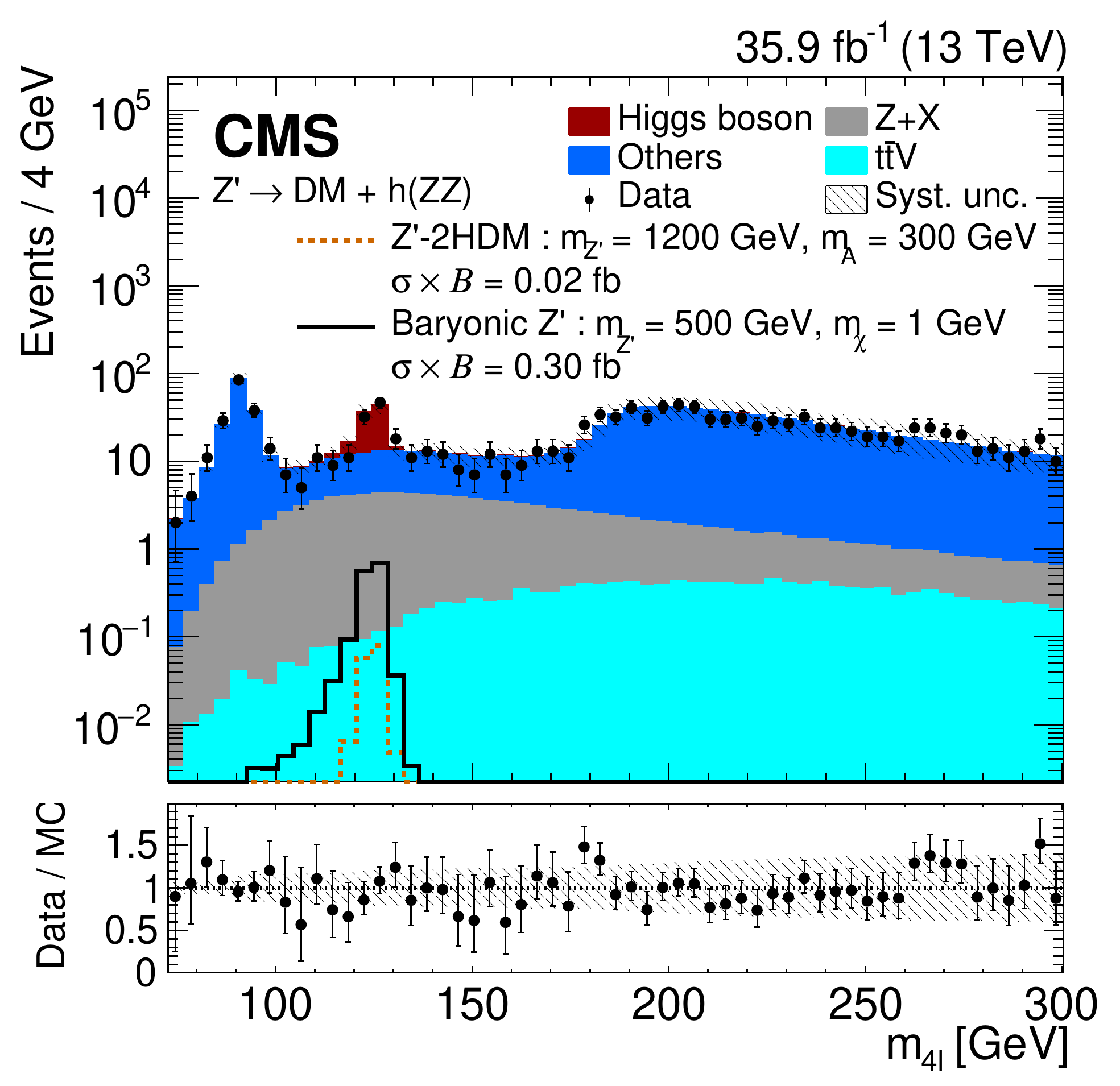}
\includegraphics[width=0.475\linewidth]{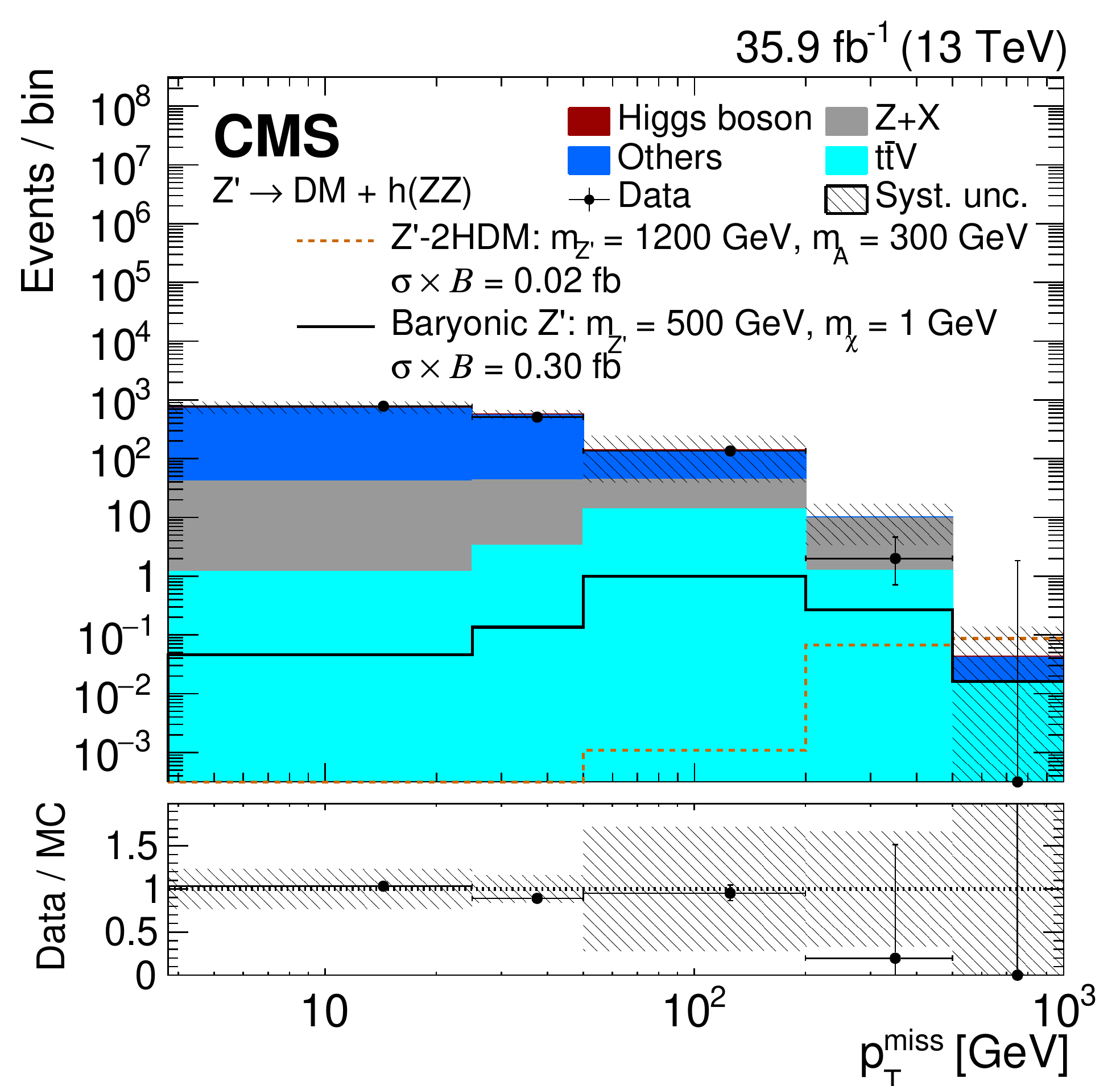}
\caption{The four-lepton invariant mass (left) and \ptmiss distributions (right) after the preselection, expected from MC simulation (stacked histograms) and observed  in data (points with vertical bars). The systematic uncertainties, discussed in Section~\protect\ref{subsec:systWW}, are shown by the hatched region. Two signal benchmarks, corresponding to the \PZp-2HDM (dotted orange line) and baryonic \PZp (solid black line) model are superimposed. The signal is normalized to the product of cross section and $\mathcal{B}$, where $\mathcal{B}$ represents the $\Ph \to \ZZ$ branching fraction. The lower panel shows the ratio of the data to the predicted SM background. }
\label{fig:observables}
\end{center}
\end{figure}

The $\mllll$ distribution
for selected \ZZ candidates exhibits a peak around 125\GeV, as expected  for both the SM Higgs boson production and signal.  However, because of the much lower cross section, the potential signal is overwhelmed by the background after the SM Higgs boson selection, as shown in Fig.~\ref{fig:observables} (left). The distribution of \ptmiss for selected \ZZ candidates is shown in Fig.~\ref{fig:observables} (right).

After the preselection, the remaining background comes from the SM Higgs boson (mostly $\PV\Ph$), $\ttbar$+\PV, and $\PV\PV/\PV\PV\PV$ production.  Another background dominated by the
$\PZ$+jets production (``$\PZ$+\PX'')~\cite{Sirunyan:2017exp} arises from secondary leptons misidentified as prompt because of the decay of heavy-flavor hadrons and light mesons within jets, and, in the case of electrons, from photon conversions or charged hadrons overlapping with photons from $\PGpz\to \PGg\PGg$ decays.  The nonprompt-lepton background also contains smaller contributions from $\ttbar$+jets, $\PZ\PGg$+jets, $\wz$+jets, and $\WW$+jets events, with a jet misidentified as a prompt lepton.  These backgrounds do not exhibit peak in the distribution of $\mllll$, and are reduced by applying a selection on the $\mllll$ around the Higgs boson mass ($115 < \mllll < 135\GeV$), by rejecting events with more than four leptons, and by requiring the number of \PQb tagged jets in the event to be less than two.

\subsubsection{Background estimation}\label{sec:background}

The dominant irreducible backgrounds from the SM Higgs boson and nonresonant \ZZ production are determined from simulation, while the $\PZ$+\PX background is determined from data~\cite{Sirunyan:2017exp}. All other backgrounds are determined from simulation. Background contributions from the SM Higgs boson production in association with a \PZ boson or a \ttbar pair, followed by the $\Ph \to \PW\PW \to 2\ell 2\nu$ decay, have been studied with simulated events and found to be negligible.

The $\PZ$+\PX background is estimated from data by first determining the lepton misidentification probability in a dedicated CR and then using it to derive the background contribution in the SR. The lepton misidentification probability is defined as the probability that a lepton passing a loose selection with relaxed identification or isolation criteria also passes the tight selection criteria.
The misidentification probability is measured in a $\PZ$+lepton CR where the \PZ boson candidate (with the mass within 7\GeV of the nominal \PZ boson mass) is formed from the two selected leptons passing the tight identification criteria, and an additional lepton is required to pass the loose selection. This sample is dominated by $\PZ$+nonprompt-lepton events. The electron and muon misidentification probabilities  are measured as functions of the lepton candidate \pt, its location in the barrel or endcap region of the ECAL or the muon system, and \ptmiss in the event, using $\PZ(\to\ell\ell)$+\Pe and $\PZ(\to \ell\ell)$+\Pgm events, respectively, in the $\PZ$+lepton CR. The misidentification probabilities are found to be independent of the charge of the lepton within the uncertainties.

The strategy for applying the lepton misidentification probabilities relies on two additional CRs. The first CR is defined by requiring that the two leptons that do not form the ${\PZ_1}$ candidate, pass only the loose, but not the tight identification criteria.
This CR defines the ``2 pass + 2 fail'' (2P2F) sample and is expected to be populated by events that intrinsically have only two prompt leptons (mostly from DY production, with a small contribution from \ttbar and $\PZ\PGg$ events). The second CR is defined by requiring only one of the four leptons to fail the tight identification and isolation criteria and defines the ``3 pass + 1 fail'' (3P1F) sample, which is expected to be populated by the type of events that populate the 2P2F CR, but with different relative proportions, as well as by $\wz$+jets events with three prompt leptons.

\section{Statistical combination of the search channels}\label{combo}

The analyses in the five channels described above are almost completely statistically independent of each other, allowing these analyses to be combined
 without accounting for the possibility  of events being selected in more than one final state. Whenever an explicit veto ensuring the strict mutual exclusivity
of the channels is not placed in a particular analysis, it was checked that there are no overlapping events with the other channels. The summary of the vetoes on additional objects, namely electrons, muons, \PGt\ leptons, photons, jets, and \PQb tagged jets, in each analysis is presented in Table~\ref{tab:vetos}. These selections not only reduce the major backgrounds, but also ensure the nearly complete mutual exclusivity of the analyses considered for the combination. The overlap in the SR is zero and for the CR it is less than 0.01\%, i.e., it is much smaller than the systematic uncertainty in the analysis. 

\begin{table*}[t]
\small
\topcaption{\label{tab:vetos}Summary of the maximum number of additional objects allowed in an event for each analysis. A dash means that no restriction on the corresponding object is applied in the corresponding analysis. }
\begin{center}
\begin{tabular}{l c c c c c }
\hline
Object  & \hbb & \hgg & \htt & \hww & \hzz \\
\hline
Electron & =0 & \NA & =0 & =0 & =0 \\
Muon      & =0 & \NA & =0 & =0 & =0 \\
\PGt\ lepton     & =0 & \NA & \NA & =0 & \NA \\
Photon       & =0 & \NA & \NA & \NA & \NA \\
AK4 Jet     & $\leq$1 & $\leq$2 & \NA & \NA & \NA \\
\PQb tagged AK4 jet   & =0 & \NA & =0 & =0 & $\leq$1 \\
\hline
\end{tabular}
\end{center}
\end{table*}

The combination of the analyses in the five Higgs boson decay channels is performed for both the \PZp-2HDM and the baryonic \PZp model. For each model, the \hbb channel dominates the sensitivity in most of the phase space, and hence the combined results are dominated by this channel. However, there are regions of the parameter space that are
hard to probe with \hbb decays, and other channels play a major role there.  The analysis strategies for all channels are the same for both models, except for the \hbb channel, where two different strategies are used because of the difference in the Lorentz boost of the Higgs boson. In this channel, the results for the \PZp-2HDM are taken from Ref.~\cite{b2g:monohbb2016}, whereas for the baryonic \PZp model, the results from Ref.~\cite{exo:monohbb2016} are used in the combination.

For the \PZp-2HDM, the two parameters that we scan are \mZp and \mA. All five analyses contribute to the combination in the ranges $800 <\mZp< 2500\GeV$ and $300 < \mA < 800\GeV$. For $\mZp < 800\GeV$, it is not possible to perform the \hbb analysis efficiently, therefore only four other decay channels are used for the combination. For $\mzp > 2500\GeV$ and $\mA >800\GeV$ the signal selection efficiency is significant only for the \hbb decay mode, hence only the \hbb channel contributes in this region.

For the baryonic \PZp model, the two parameters that we scan are \mZp and \mdm, and all five analyses are performed in the full phase space considered for the combination. Since the maximum sensitivity for all the analyses is achieved for $\mdm = 1\GeV$, the comparison of individual analyses is shown only for this DM particle mass, to demonstrate the improvement in the sensitivity achieved in the combination of individual channels.

\section{Systematic uncertainties} \label{sec:systematics}

A number of systematic uncertainties are considered in the combination, broadly divided into two categories: theoretical and experimental. Theoretical uncertainties are considered fully correlated
among all five channels. Only the systematic uncertainties attributed to the experimental sources that are correlated between different channels are described for the combined result in section~\ref{subsec:systcombo}. The details of all experimental systematic uncertainties in the \hbb analysis using AK8 jets are described in Ref.~\cite{b2g:monohbb2016} and those for the analysis using CA15 jets are described in Ref.~\cite{exo:monohbb2016}; for the \hgg and \htt channels they are given in Ref.~\cite{exo:monohggtt2016}; and for the \hww and \hzz analyses they are discussed in this section.

\subsection{The \texorpdfstring{$\hwwp+\ptmiss$}{h(WW) + pTmiss} channel} \label{subsec:systWW}

The normalization and the kinematic shapes of the BDT discriminant distributions for the main backgrounds are derived from data CRs, and therefore systematic uncertainties in both the normalization and shapes are considered.

For the nonprompt-lepton background the uncertainty amounts to approximately 30\%, and covers the uncertainty in the lepton misidentification rate, the dependence on the CR background composition, and the statistical component because of the finite event count in the CR.

The top quark background CR is included as an additional category in the signal extraction fit. The kinematic shapes of the top quark background are taken from simulation corrected for the \PQb tagging scale factors, with the uncertainties covering the difference between the \PQb tagging efficiency in data and simulation~\cite{Sirunyan:2017ezt}. A similar procedure is applied for the DY background, by defining a CR in low-\mT phase space, and to the nonresonant \WW background, for which a high-\mll CR is defined.
The top quark and DY background normalizations are correlated between their respective CRs and the SR and are left unconstrained in the fit.
The change in the PDF, the renormalization and the factorization scale variations from their nominal values lead to migration of the top quark and \dytt background
events between the respective CRs and the SR. To take into account this effect, the change in the top quark ( \dytt) background yield is used as an additional
1 (2)\% uncertainty in the corresponding CR.
The shapes of the \wz and other minor backgrounds are taken from simulation and normalized to their theoretical predictions, with the theoretical uncertainties estimated. The uncertainties related to the modeling of \ptmiss are estimated by considering the effect of varying the lepton energy scale, JES, JER, and unclustered energy scale on \ptmiss.

Experimental uncertainties are estimated by applying scale factors between data and simulation, and/or by smearing of certain kinematic variables in simulation, with the corresponding changes further propagated to all analysis variables. The signal acceptance uncertainty associated with the combination of single-lepton and dilepton triggers is measured to be 2\%.
The uncertainty in the ratio between the single top quark and top quark pair production cross sections, 8\% at 13\TeV~\cite{Chatrchyan:2013iaa}, has been also included, as it affects the top quark background yield from the maximum-likelihood fit used to extract the signal and dominant backgrounds. The uncertainty in the \pt spectrum of the top quark has been applied to all the observables in order to cover the difference between the simulated and observed spectra~\cite{Sirunyan:2018ptc}, and is of the order of 1\%.

The uncertainty in the Higgs boson branching fraction for the \hww decay is about 1\%~\cite{yr4}.
The uncertainty in the NNLO $K$ factor applied to the LO $\Pg\Pg \to \PW\PW$ cross section estimate is 15\%~\cite{Passarino:2013bha}. The \ptww spectrum in the $\PQq\PQq \to \PW\PW$ sample has been reweighted to match the resummed calculation~\cite{Meade:2014fc,Jaiswal:2014yba}. The associated shape uncertainties related to the missing higher-order corrections are modeled by varying the factorization, renormalization, and resummation scales up and down independently by a factor of 2 from their nominal values~\cite{Meade:2014fc}. Finally, uncertainties arising from the limited size of the simulated samples are included for each bin of the BDT discriminant distributions, in each category. The main sources of the uncertainties affecting the analysis are  listed in Table~\ref{tab:HwwSystematics}.

\begin{table}[htbp]
  \centering
  \caption{
    Systematic uncertainties affecting the \hww analysis.
  \label{tab:HwwSystematics}}
  \begin{tabular}{lcc}
    \hline
    Source of uncertainty                     & Process & Size      \\
    \hline
    Trigger efficiency                        & Simulated samples  & 2\%         \\
    Nonprompt lepton bkg.             & Nonprompt lepton bkg. & 30\%        \\
    \wz bkg. normalization                          & $\PW\PZ$      & 16\%        \\
    $\PW\PGg^{(*)}$ bkg. normalization                 & $\PW\PGg^{(*)}$      & 26\%        \\[\cmsTabSkip]
    \hww branching fraction                  & Signal        & $\sim$1\%  \\
    Single \PQt/\ttbar cross section ratio  & Top quark               & 8\%         \\
    Top quark \pt                             & Top quark               & 1\%         \\
    \ggww LO to NNLO $K$ factor                & \ggww              & 15\%        \\
    \ptww resummation                        & \qqww             & $\sim$5\%    \\
    Top quark CR to SR transfer factor              & Top quark                & 1\%         \\
    \dytt CR to SR transfer factor           & \dytt              & 2\%         \\
    Simulated sample event count & Simulated samples & 2--70\% \\
    \hline
  \end{tabular}
\end{table}

\subsection{The \texorpdfstring{$\hzzp+\ptmiss$}{h(ZZ) + pTmiss} channel}
A source of systematic uncertainty in the nonprompt-lepton background estimate potentially arises from the difference in the composition of the SM background processes with nonprompt leptons
($\PZ\PGg$+jets, \ttbar, $\PZ\PGg$+jets) contributing to the CRs where the lepton misidentification rate is measured and applied.
This uncertainty can be estimated by measuring the misidentification rates in simulation for the 2P2F and 3P1F CRs. Half of the difference between the misidentification rates obtained from simulation in these two CRs is used as a measure of the systematic uncertainty in the lepton misidentification rate and is further propagated to the uncertainty in the nonprompt-lepton background, and amounts to 43\% for the $4\Pe$, 36\% for the $4\Pgm$, and 40\% for the $2\Pe2\Pgm$ final states.

 The uncertainty in the full signal selection efficiency is at the level of 1\%.  The uncertainty in the $\mllll$  resolution from the uncertainty in the per-lepton energy resolution is about 20\%~\cite{Sirunyan:2017exp} and affects the signal and all the backgrounds from Higgs boson production.

In addition, there are two types of systematic uncertainties related to the
modeling of $\ptmiss$. The first uncertainty is related to the approximately Gaussian core of
the resolution function for correctly measured jets and other physics objects and corresponds to the
uncertainty in the genuine $\ptmiss$. The second uncertainty, attributed to
significant mismeasurement of \ptmiss, is an uncertainty in the  ``mismeasured'' $\ptmiss$.

The uncertainties from the modeling of genuine $\ptmiss$ are measured by varying the parameters
associated with the corrections applied to $\ptmiss$ and by propagating those variations to the $\ptmiss$ calculation,
after applying the full analysis selection. Each correction is varied up and down by one standard deviation of the
input distribution. The corrections used in this  calculation come from JES, JER,
muon, electron, photon, and the unclustered energy scales.

The uncertainty in the  mismeasured $\ptmiss$ is obtained from a sample with significant
contributions from misidentified leptons and mismeasured jets, obtained by requiring
an opposite-sign, same-flavor dilepton pair passing the $\PZ_{1}$
candidate selection, and an additional same-sign, same-flavor pair (``OS+SS'' sample). This sample is enriched in misidentified leptons that form the same-sign pair and is expected to lead to significant mismeasurement of $\ptmiss$, not
already covered by the uncertainties in the Gaussian core discussed above. We derive the mismeasured $\ptmiss$ uncertainty from the comparison of the $\ptmiss$ shapes in the ``OS+SS'' sample and in the SR, with a requirement that the $\mllll$ be outside the Higgs boson invariant mass peak ($\abs{\mllll - 125\GeV} > 10\GeV$). The uncertainty in mismeasured \ptmiss is applied to the $\PZ$+\PX sample only, since the effect is expected to be negligible when four genuine leptons are produced, as is the case for the signal and for most of the simulated background samples.

An uncertainty of 10\% in the $K$ factor used for the \ggZZ\ prediction is applied~\cite{Passarino:2013bha}. A systematic uncertainty of 2\% in the \HZZfl branching fraction~\cite{yr4} affects both signal and the SM Higgs boson background yields. Theoretical uncertainties in the $\ttbar$+\PV background cross sections are taken from Ref.~\cite{Frixione2015}. A summary of the experimental uncertainties is given in Table~\ref{tab:SystOverview}.

\begin{table}[!htb]
\begin{center}
\caption{Systematic uncertainties affecting the \hzz analysis. }
\label{tab:SystOverview}
\begin{tabular}{lcc}
\hline
Source of uncertainty & Process & Rate \\
\hline
Trigger selection & Simulated samples & 2\% \\
$\mllll$ resolution & Higgs boson &  20\% \\
\HZZfl branching fraction & Higgs boson & 2\% \\
 \ggZZ NNLO to LO $K$ factor & \ggZZ\ bkg. & 10\% \\
Genuine \ptmiss & Simulated samples (Shape) &  7--26\% \\
Mismeasured \ptmiss & $\PZ$+\PX bkg. (Shape) & 2--30\% \\
$\PZ$+\PX bkg. yield & $\PZ$+\PX bkg. (Yield) & 36--43\% \\
$\ttbar$+\PV bkg. yield & $\ttbar$+\PV bkg. & 27--34\% \\
\hline
\end{tabular}
\end{center}
\end{table}

\subsection{Systematic uncertainties in the combination}\label{subsec:systcombo}

\begin{table}
\caption{Systematic uncertainties in the combination of channels, along with the type (rate/shape) of uncertainty affecting signal and background processes, correlated amongst at least two final states. For the rate uncertainties, the percentage of the prior value is quoted, while for shape uncertainties an estimate of the impact of systematic uncertainties on the yield is also listed.  A dash (``\NA'') implies that a given uncertainty does not affect the analysis. Whenever an uncertainty is present but kept uncorrelated in a particular channel, this is mentioned explicitly. The effect of the \PQb jet mistag rate uncertainty is very small in the \hbb Z'-2HDM analysis and hence it is added to the effect of the \PQb tagging efficiency uncertainty in quadrature.}
\begin{center}
  \cmsTable{
\begin{tabular}{ lcccccc }
\hline
Source &  \multicolumn{2}{c}{\hbb} & \hgg & \htt & \hww & \hzz  \\
                             &  \PZp-2HDM & Baryonic \PZp          &                         &      &      &       \\
\hline
AK4 jet \PQb tagging             &          &  Uncorr. (3--4\%) &   \NA    &   4\%                        & Shape (1\%) &  1\%      \\  [-2.3ex]
AK4 jet \PQb mistag              &    \raisebox{1.5ex}{{\huge{\}}}3--11\%}                     & Shape (5--7\%) &   \NA    &  2--5\%  & Shape (1\%) &  \NA     \\
$\Pe$ ident. efficiency          &    4\%      &  2\%           &   \NA    &   2\%           & Shape (2\%) &  2.5--9.0\% \\
\PGm ident. efficiency                &    4\%      &  2\%           &   \NA    &   2\%      & Shape (2\%) &  2.5--9.0\% \\
\tauh ident. efficiency             &    3\%      &  3\%           &   \NA    &  4.5\%       & Shape (1\%) &  \NA     \\
$\Pe$ energy scale        &    1\%      &  \NA           &   \NA    &  \NA                   & Shape (1\%) &  3\%     \\
$\PGm$ energy scale              &    1\%      &  \NA           &   \NA    &  \NA            & Shape (1\%) &  0.4\%    \\
JES             &    \NA      &  Uncorr. (4\%) &   \NA                        & Shape ($<$10\%) & Shape (3\%) &  2--3\%  \\
Int. luminosity                   &    2.5\%    &  2.5\%         &  2.5\%   &  2.5\%  &  2.5\%  &  2.5\%   \\
Signal (PDF, scales) 		 &    0.3--9.0\%    &  0.3--9.0\%          &  0.3--9.0\%    &  0.3--9.0\%   &  0.3--9.0\%   &  0.3--9.0\%    \\
\hline
\end{tabular}
}
\end{center}
\label{table:nuisancescombo}
\end{table}

The uncertainties associated with the background normalization and fit parameters are assumed to be uncorrelated, whereas those associated with the standard object selection are considered fully correlated and are summarized in Table~\ref{table:nuisancescombo}. In all five decay channels, a normalization uncertainty of 2.5\% for simulated samples is used to account for the uncertainty in the measurement of the integrated luminosity~\cite{CMS-PAS-LUM-17-001}. Also fully correlated across all channels are the systematic uncertainties related to theoretical calculations of the Higgs boson production cross section, PDFs, and renormalization and factorization scale uncertainties estimated using the recommendations of the PDF4LHC~\cite{Butterworth:2015oua} and LHC Higgs Cross Section~\cite{yr4} working groups, respectively. These uncertainties range from 0.3 to 9.0\%.

Uncertainties from imprecise knowledge of the JES are evaluated by propagating the uncertainties in the JES for individual jets in an event, which depend on the jet \pt and $\eta$, to all the analysis quantities. The uncertainties in the selection of \PQb tagged AK4 jets are taken into account using the uncertainties in the \PQb tagging efficiency and misidentification rate estimated from the difference
 between data and simulation~\cite{Sirunyan:2017ezt}. The uncertainty due to the difference in the performance of electron, muon, and $\PGt$ lepton identification between data and simulation is taken into account for individual decay channels and considered fully correlated in the statistical combination. An uncertainty of 1--3\% in the electron energy scale and an uncertainty of 0.4--1.0\%  in the muon energy scale are considered to be correlated in the combination.

\section {Results} \label{sec:results}

The event selection described in Section \ref{sec:eventselection} has been used to discriminate the mono-Higgs signal from backgrounds in each channel. The observed yields in data and the expected event yields for the signal and background processes in the \hbb, \hgg, and \htt channels can be found in Refs. \cite{b2g:monohbb2016,exo:monohbb2016,exo:monohggtt2016}. The corresponding yields for the \hww and \hzz analyses are discussed in Section~\ref{results:hWW-hZZ}.
Tables \ref{tab:MonoHiggsWW_yields}, \ref{tab:MonoHiggs_yields} and figures \ref{fig:WWMVA_full_selection}, \ref{fig:ZZMET_full_selection} show one signal mass hypothesis for each model, normalized to the respective cross section.
For the \PZp-2HDM, the signal is normalized to the cross section calculated for  mass values of \PZp and \PA bosons of 1200 and 300\GeV, respectively, and for $\gZp = 0.8$, $\tan{\beta} = 1$.
For the baryonic \PZp model, the signal is normalized to the cross section corresponding to the \PZp and \mdm masses of 500 and 1000\GeV, respectively, and for $g_{\PGc} = 1$, $\gq = 0.25$.
\subsection{The \texorpdfstring{$\hwwp+\ptmiss$}{h(WW) + pTmiss} and \texorpdfstring{$\hzzp+\ptmiss$}{h(ZZ) + pTmiss} channels}\label{results:hWW-hZZ}

The expected background yields and the observed number of event in data, along with the expected yields for two signal benchmarks in the \hww and \hzz channels, are summarized in Tables~\ref{tab:MonoHiggsWW_yields} and \ref{tab:MonoHiggs_yields}, respectively.

\begin{table*}[t]
\caption{The post-fit signal and background event yields, and the observed number of events in data, for the \hww analysis. The expected numbers of signal events for the two signal hypotheses are also reported, one for each benchmark model. The total uncertainty, including both statistical and systematic components, is quoted for the expected signal and backgrounds yields.}
\begin{center}
\begin{tabular}{lcc}
  \hline
Channel                                                     & Event yield       \\
  \hline
  SM Higgs boson ($m_{\Ph} = 125\GeV$)         & $598   \pm 55$\\
  Top quark                                                                 & $4\,450  \pm 310$ \\
  \WW                                                                   & $4\,500 \pm 160$ \\
  Other $\PV\PV/\PV\PV\PV$                                                    & $449  \pm 44$ \\
  $\PZ$+jets                                                             & $367   \pm 42$  \\
  Nonprompt lepton bkg.                                                    & $660 \pm 210$ \\[\cmsTabSkip]
  Total bkg.                                                  & $11\,030 \pm 410$ \\[\cmsTabSkip]
  \PZp-2HDM ($\mZp = 1200\GeV$, $\mA = 300\GeV$)     & $3.04   \pm 0.10$   \\
 Baryonic \PZp ($\mZp = 500\GeV$, $\mdm = 1\GeV$) & $29.60 \pm 0.89$  \\ [\cmsTabSkip]
  Observed                                                             & 11\,172      \\ \hline
\end{tabular}
\end{center}
\label{tab:MonoHiggsWW_yields}
\end{table*}

\begin{table*}[t]
\caption{The post-fit signal and background event yields, and the observed number of events in data, for the \hzz analysis. The expected numbers of signal events for the two signal hypotheses are also reported, one for each benchmark model. The total uncertainty, including both statistical and systematic components, is quoted for the expected signal and backgrounds yields.}
\begin{center}
\cmsTable{
\begin{tabular}{lcccc}
\hline
Channel & $4\Pe$ & $4\PGm$ & $2\Pe2\PGm$ & $4\ell$ \\
\hline
SM Higgs boson ($m_{\Ph} = 125\GeV$) & $12.1 \pm 1.4$ & $21.1 \pm 1.9$ & $27.9 \pm 2.4$  & $61.1 \pm 4.8$ \\
$\PZ\PGg^*$, \ZZ & $7.0^{+0.9}_{-1.2}$ & $14.7^{+1.1}_{-1.2} $  & $18.4^{+1.7}_{-1.8}$  & $40.1^{+3.2}_{-3.6}$ \\
$\ttbar$V & $0.10 \pm 0.02 $ & $0.07 \pm 0.02$ & $0.12 \pm 0.02$  & $0.29 \pm 0.05$  \\
$\PV\PV\PV$ & $0.04 \pm 0.03$ & \NA & $0.03 \pm 0.03$  & $0.07 \pm 0.06$ \\
$\PZ$+\PX & $3.0 \pm 2.1$ & $4.7 \pm 2.7$ & $8.5 \pm 3.8$ & $16.2 \pm 4.9$ \\[\cmsTabSkip]
Total bkg. & $22.2^{+2.6}_{-2.8}$ & $40.6 \pm 3.8$ & $55.0 \pm 4.8$  & $117.8^{+7.5}_{-7.7} $ \\[\cmsTabSkip]
\PZp-2HDM ($\mZp=1200\GeV$, $\mA=300\GeV$)  & $0.07 \pm 0.02$ & $0.11 \pm 0.02$ & $0.17 \pm 0.03$  & $0.36 \pm 0.06$ \\
Baryonic \PZp ($\mZp=500\GeV$, $\mdm=1\GeV$)  & $0.25 \pm 0.06$ & $0.45 \pm 0.09$ & $0.67 \pm 0.14$  & $1.38 \pm 0.25$ \\[\cmsTabSkip]
Observed & 24 & 44 & 44 & 112 \\
\hline
\end{tabular}
}
\label{tab:MonoHiggs_yields}
\end{center}
\end{table*}

\begin{figure}[htbp]
\begin{center}
\includegraphics[width=0.475\textwidth]{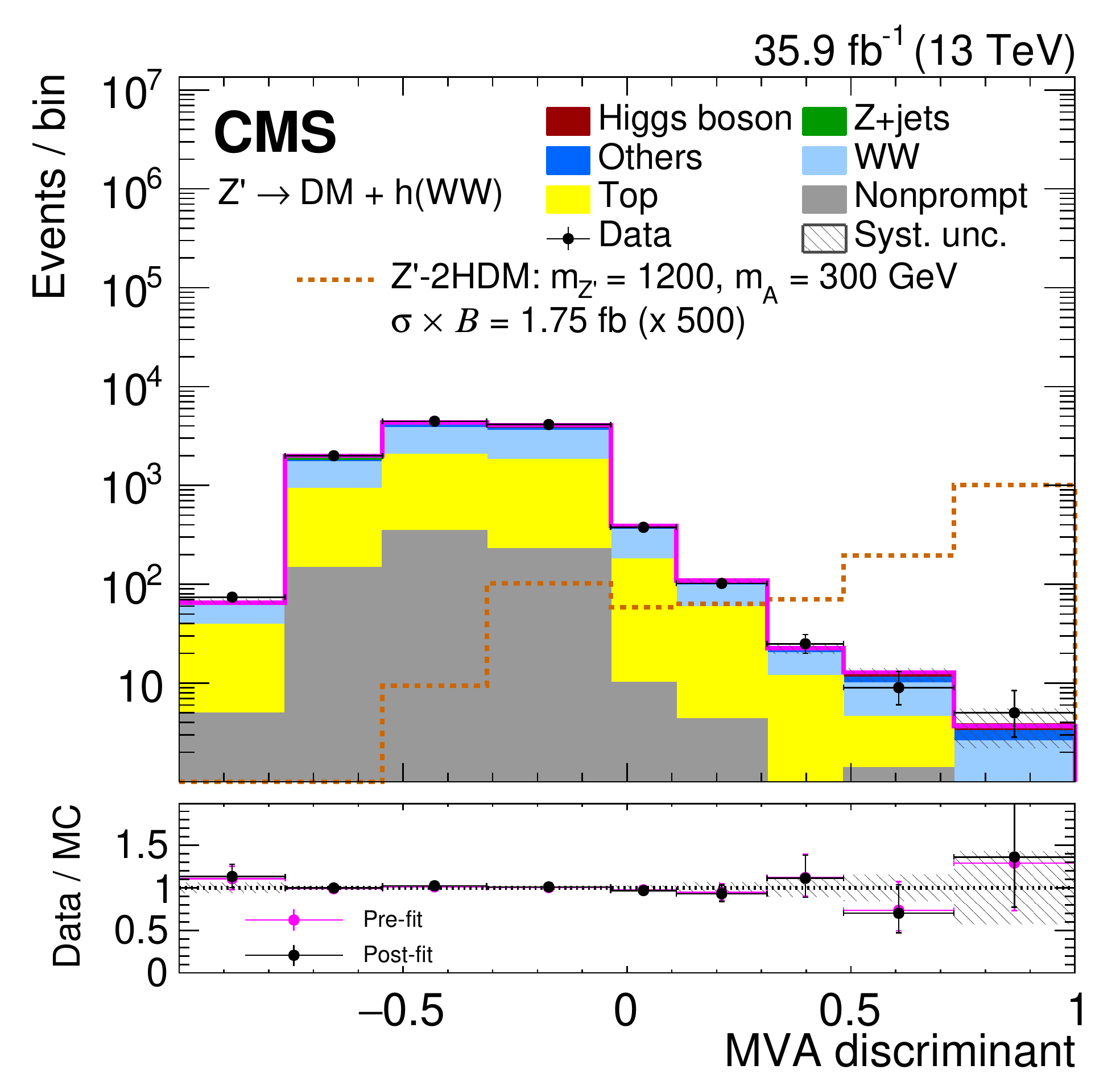}
\includegraphics[width=0.475\textwidth]{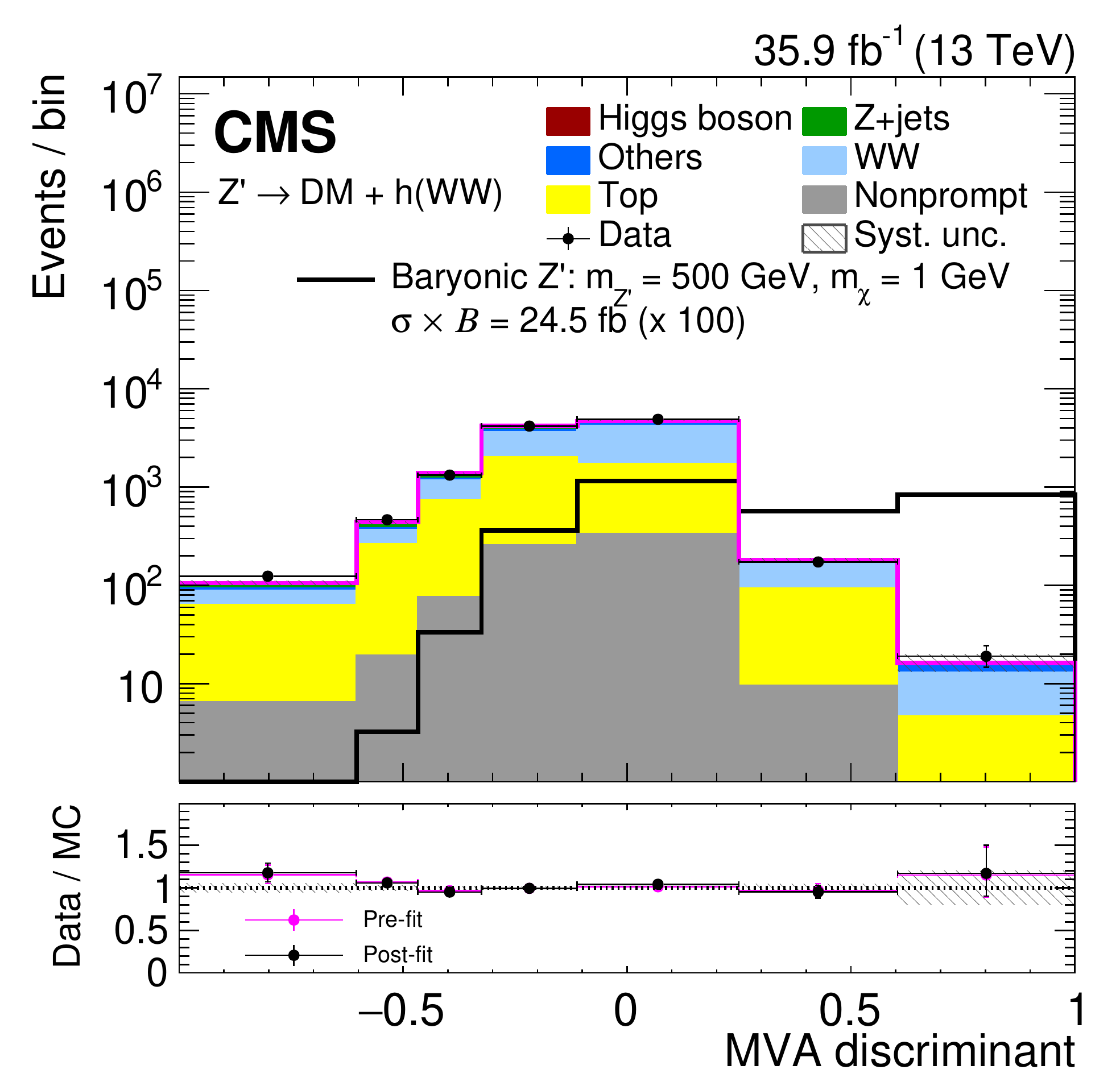}
\caption{  The distribution of the BDT discriminants expected from MC simulation before and after the fit, and observed in data (points with error bars) for the \PZp-2HDM (left) and baryonic \PZp (right) model in the signal region in the \hww analysis. Two signal benchmarks, corresponding to the \PZp-2HDM (dotted orange line, left) and baryonic \PZp (solid black line, right) model are superimposed. The signal is normalized to the product of cross section and $\mathcal{B}$ , where $\mathcal{B}$ represents the $\Ph \to \WW$ branching fraction. The signal distributions are scaled up by a factor 500 (100) for the \PZp-2HDM (baryonic \PZp model), to make them more visible. The systematic uncertainties  are shown by the hatched band. The lower panel shows the ratio of data to the total background yield, before and after the fit.}
\label{fig:WWMVA_full_selection}
\end{center}
\end{figure}

\begin{figure}[!htb]
\begin{center}
\includegraphics[width=0.49\textwidth]{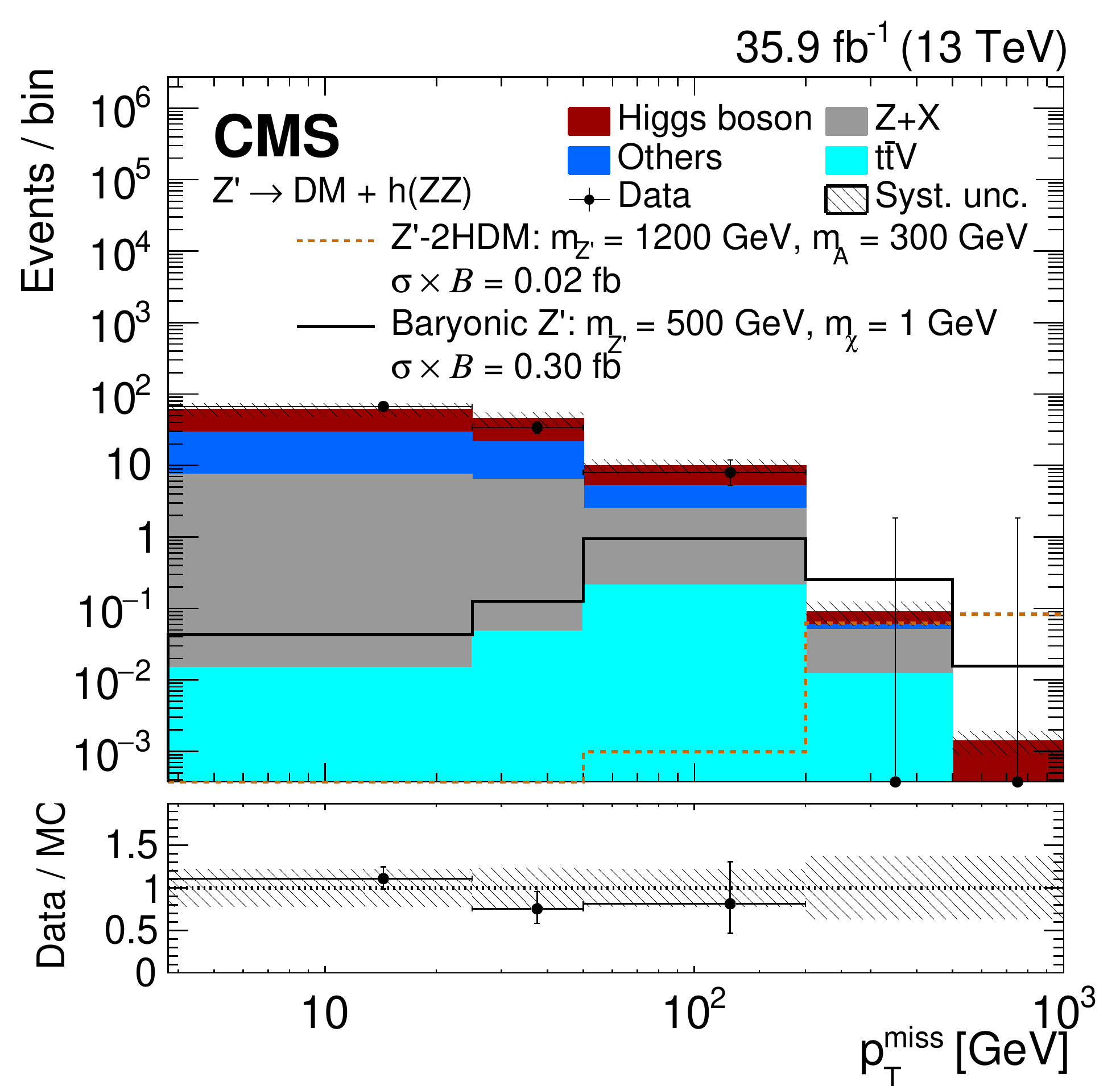}
\caption{The \ptmiss distribution for the expected background and observed events in data in the \hzz analysis. Two signal benchmarks, corresponding to the \PZp-2HDM (dotted orange line, left) and baryonic \PZp (solid black line, right) model are superimposed. The signal is normalized to the product of cross section and $\mathcal{B}$, where $\mathcal{B}$ represents the $\Ph \to \ZZ$ branching fraction. The systematic uncertainties  are shown by the hatched band. The ratios of the data and the sum of all the SM backgrounds are shown in the bottom panels. }
\label{fig:ZZMET_full_selection}
\end{center}
\end{figure}

Figure \ref{fig:WWMVA_full_selection} shows the BDT discriminant distribution for the expected backgrounds and observed events in data for the \hww analysis. Benchmark signal contributions in the  \PZp-2HDM (left) and baryonic \PZp (right) model are also shown, scaled by the factors of 500 and 100, respectively, for better visibility.
Figure~\ref{fig:ZZMET_full_selection} shows the \ptmiss distribution  of the expected backgrounds and observed events in data for the \hzz analysis. Benchmark signal contributions are also shown. For both analyses, the total uncertainty, given by a quadratic sum of the statistical and systematic components, is shown. The bottom panels show the ratios of data to the total background prediction with their total uncertainties.

The potential signal is extracted from the fit to the BDT discriminant (\ptmiss) spectrum with a signal-plus-background hypothesis for the \hww (\hzz) channel. The profile likelihood ratio is used as a test statistic, in an asymptotic approximation~\cite{Cowan:2010js}. Data agree well with the expected background and no signal is observed in either channel. Limits on the model parameters at 95\% confidence level (\CL) are set using the modified frequentist \CLs criterion~\cite{Junk,Read,combineProcedure} with all the nuisance parameters profiled.

\begin{figure}[!htb]
  \centering
    \includegraphics[width=0.49\textwidth]{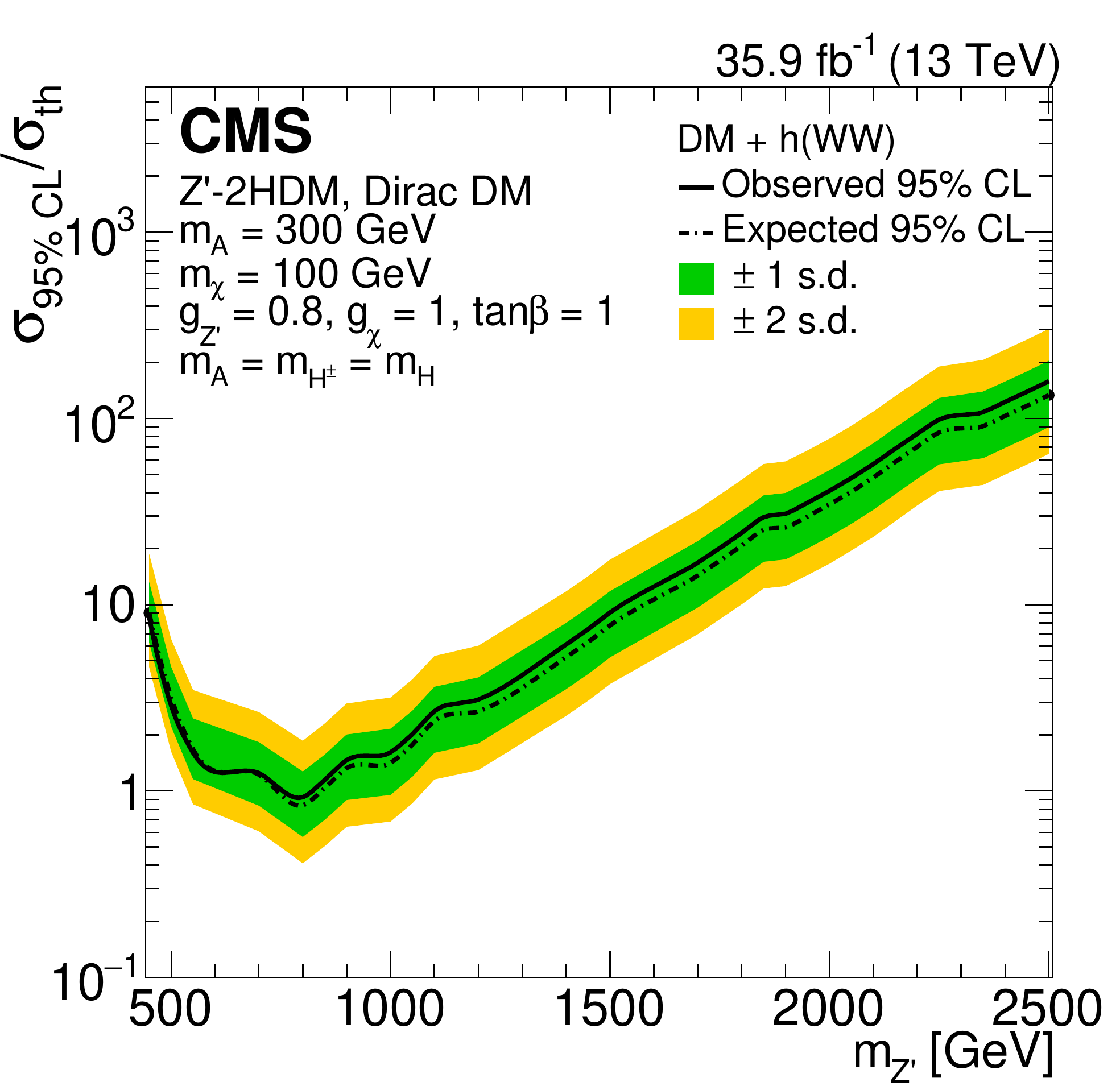}
    \includegraphics[width=0.49\textwidth]{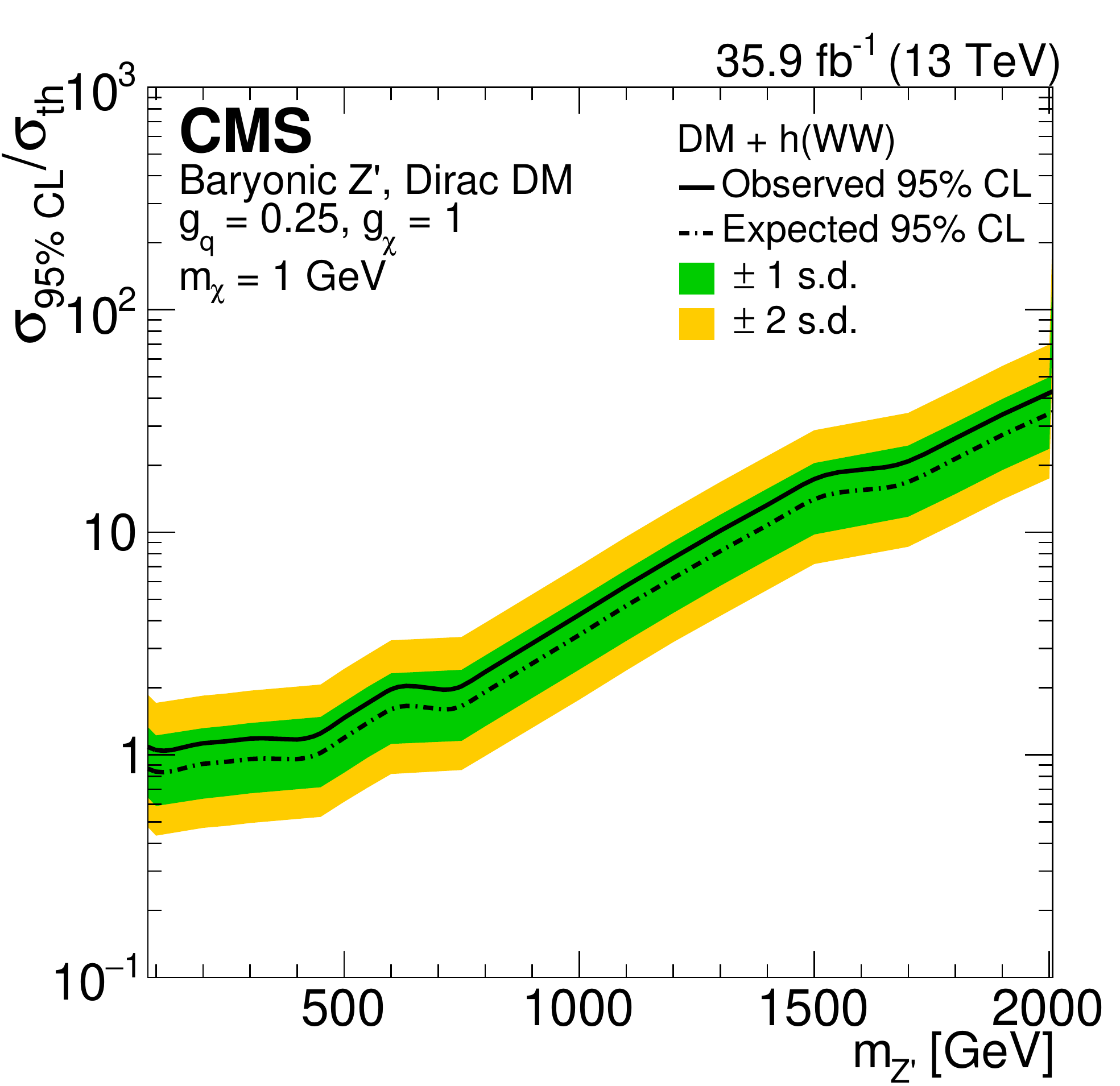} \\
    \includegraphics[width=0.49\textwidth]{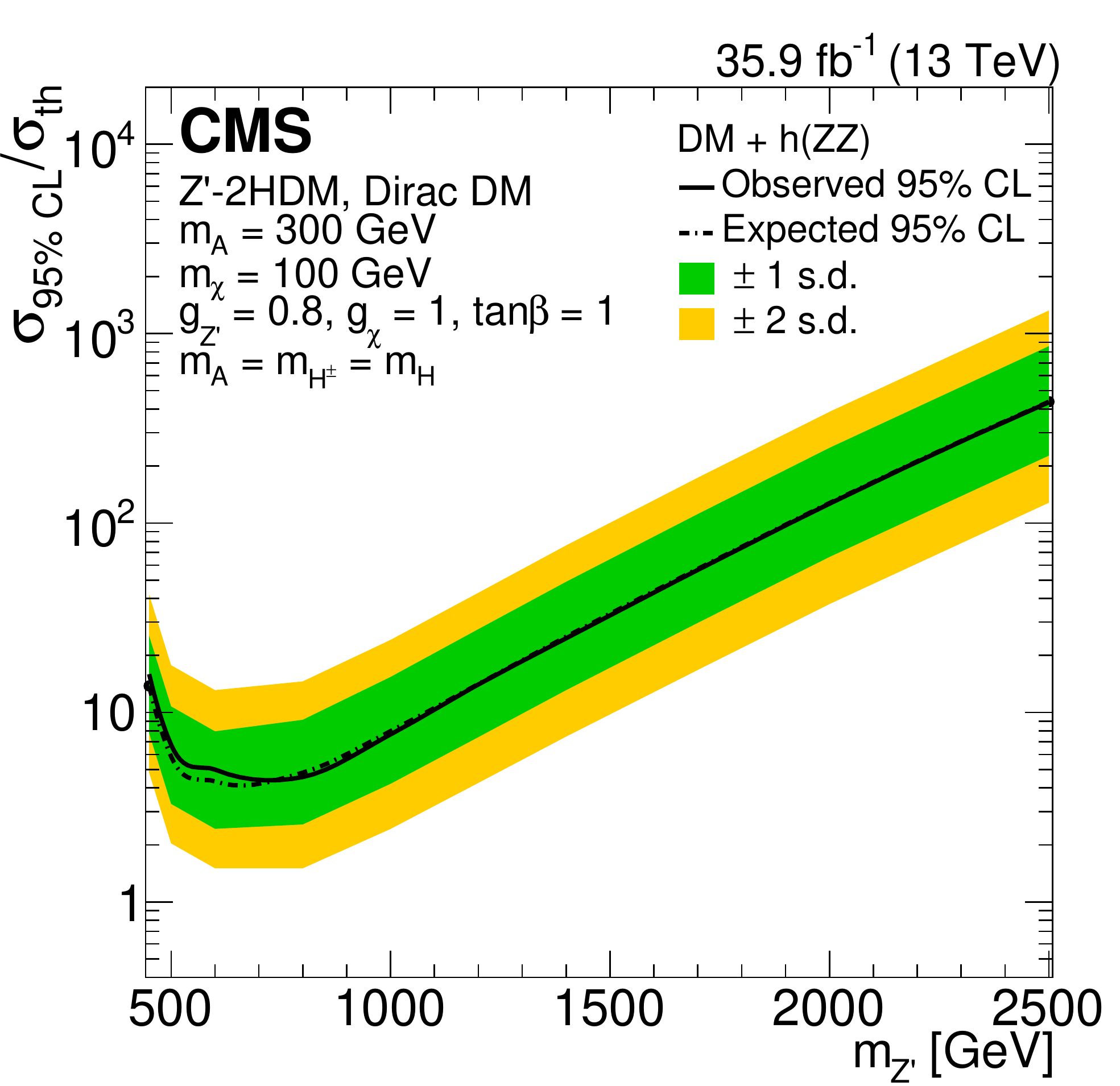}
    \includegraphics[width=0.49\textwidth]{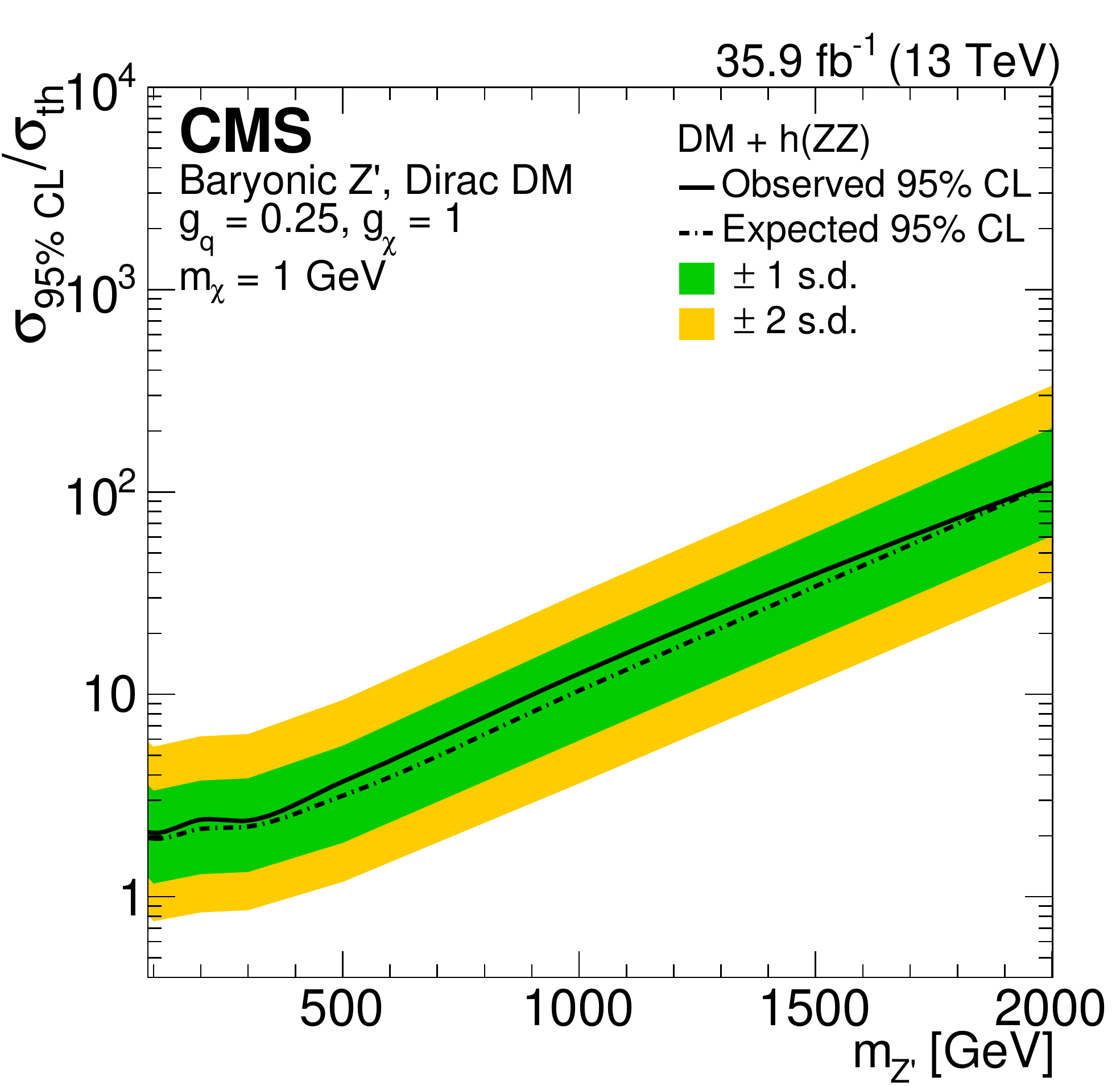}
  \caption{    The upper limits at 95\% \CL on the observed and expected DM production cross section for the \hww (upper) and \hzz (lower) analyses for the \PZp-2HDM with  $\mA = 300\GeV$ (left)  and for the baryonic \PZp with $\mChi = 1\GeV$ (right) model. The inner and outer shaded bands show the 68 and 95\% uncertainties in the expected limit, respectively.}
  \label{fig:monohwwzzlimits}
\end{figure}

The observed and expected upper limits on the DM candidate production cross section are shown in Fig.~\ref{fig:monohwwzzlimits} for the \hww (upper) and \hzz (lower) channels for the \PZp-2HDM with $\mA = 300\GeV$ (left) and for the baryonic \PZp model with the value of $m_{\chi}$ fixed at 1\GeV (right). All other model parameters are fixed to the values described in Section~\ref{sec:introduction}. The upper limits for the \hzz analysis already include the statistical combination of all three final states used.
The \hww analysis excluded the region from 780 to 830\GeV for $\mA = 300\GeV$ in the \PZp-2HDM.

\subsection{Results of the statistical combination}

\begin{figure}[htbp]
  \centering
    \includegraphics[width=0.495\textwidth]{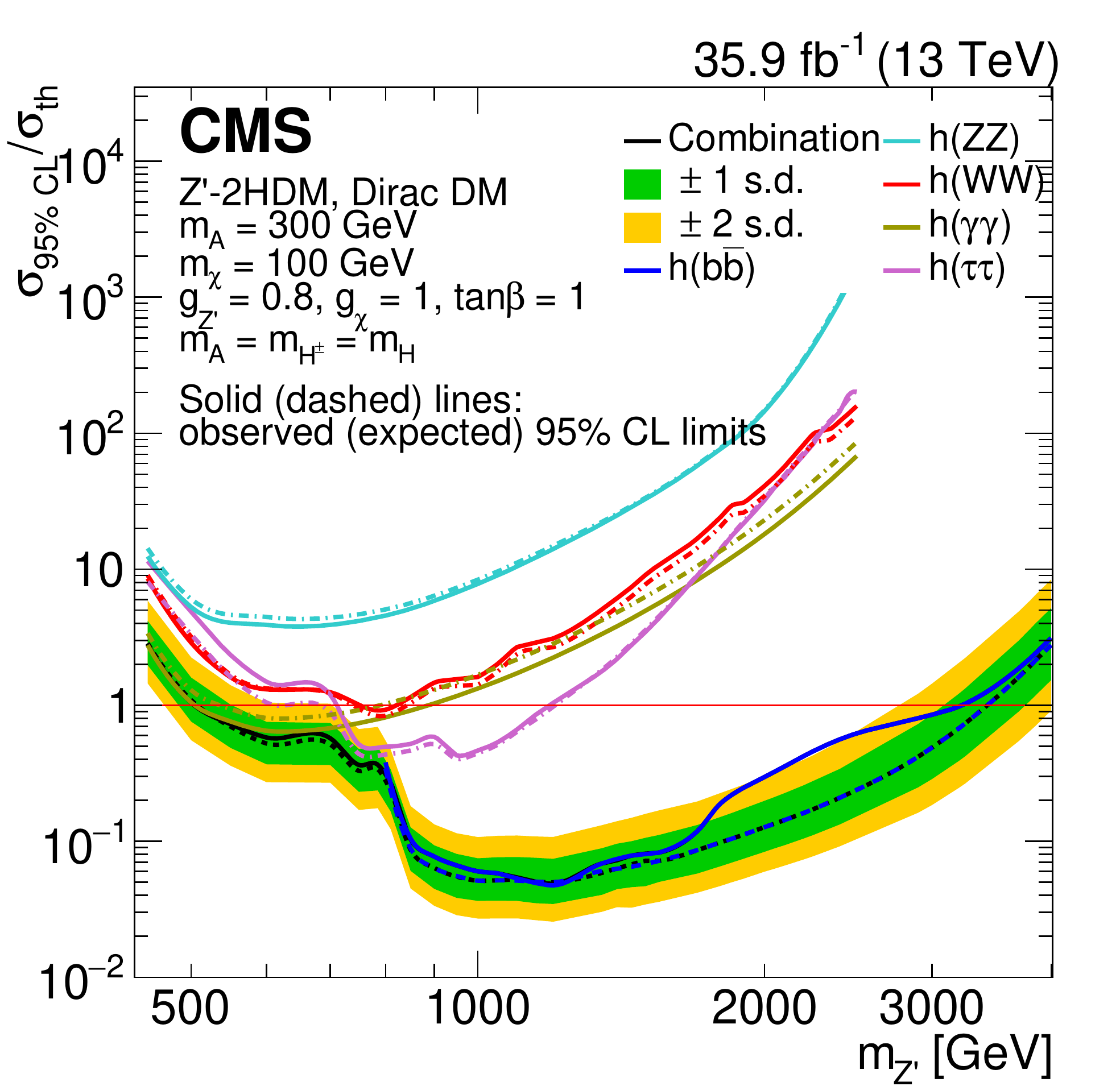}
    \includegraphics[width=0.495\textwidth]{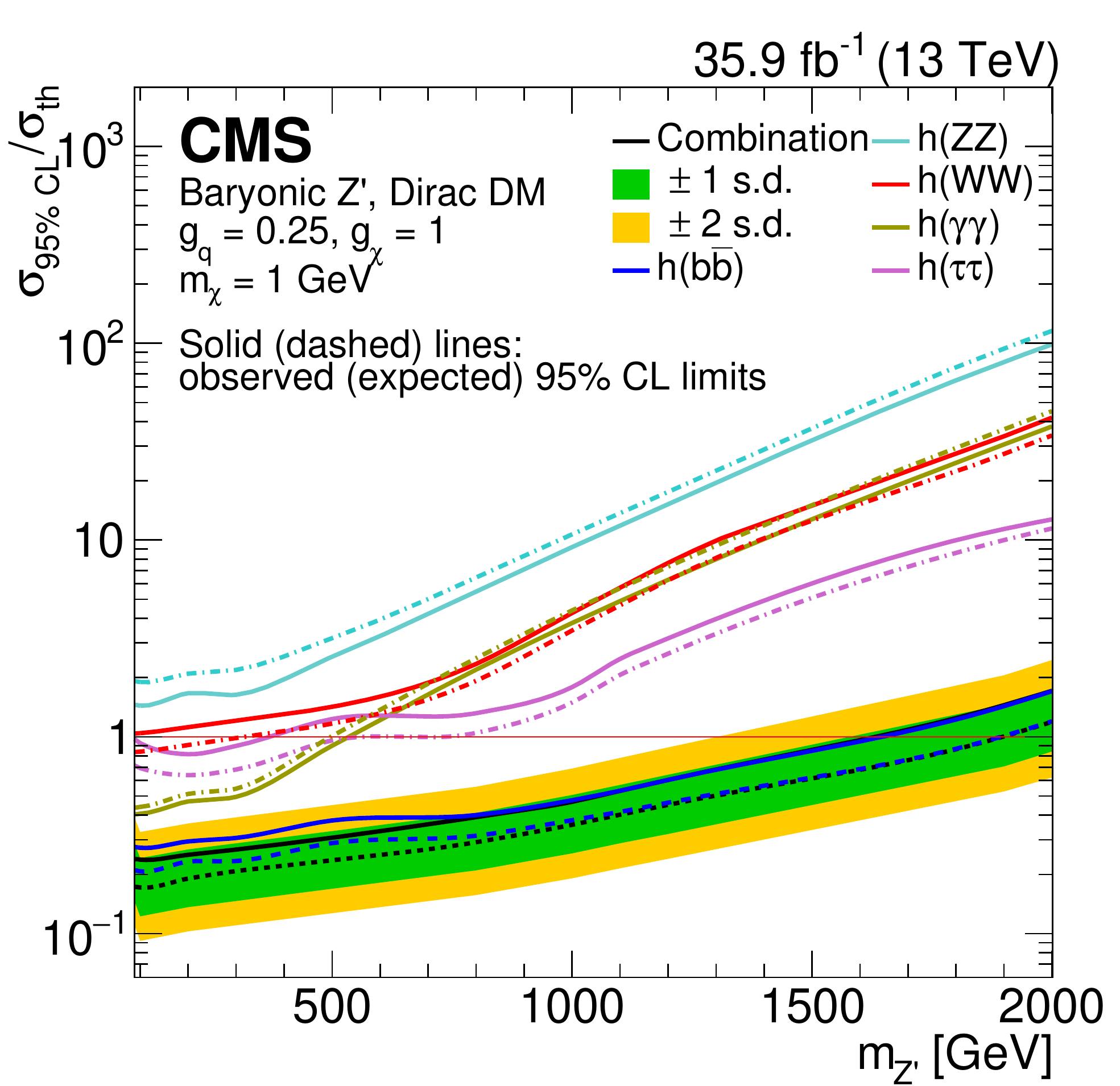}
  \caption{    The upper limits at 95\% \CL on the observed and expected $\sigma/\sigma_\text{th}$ for the \PZp-2HDM (left) and baryonic \PZp (right) model for the five individual decay modes of the Higgs boson, and for their combination. The distributions are shown as a function of \mZp for  $\mA = 300\GeV$ (\PZp-2HDM) and $\mchi = 1\GeV$ (baryonic \PZp model). The inner and outer shaded bands show the 68 and 95\% \CL uncertainties in the expected limit, respectively.}
  \label{fig:monohcombo1d}
\end{figure}

The observed and expected upper limits at 95\% \CL on the DM production cross section normalized to the predicted cross section, as a function of \mZp, from the combination of all five channels are shown in Fig.~\ref{fig:monohcombo1d} for the \PZp-2HDM with $\mA = 300\GeV$ (left) and for the baryonic \PZp model with $\mChi = 1\GeV$ (right). The combined result is also compared with those of the individual analyses.

For the \PZp-2HDM,  the combination is dominated by the \hbb analysis for $\mZp > 800\GeV$. However, the \hbb analysis has no sensitivity for \mZp values below 800\GeV, and a combination of the \hgg and \htt channels plays a significant role in this region of the model parameter space. The range of \mZp excluded at 95\% \CL spans from 500 to 3200\GeV for $\mA = 300\GeV$.

For the baryonic \PZp model, the combination results are also dominated by the \hbb channel, but the \hgg and \htt channels also provide a nonnegligible contribution in constraining the model parameters.
The range of \mZp excluded at 95\% \CL spans from 100 to 1600\GeV for  $\mChi = 1\GeV$.

\begin{figure}[htbp]
  \centering
    \includegraphics[width=0.495\textwidth]{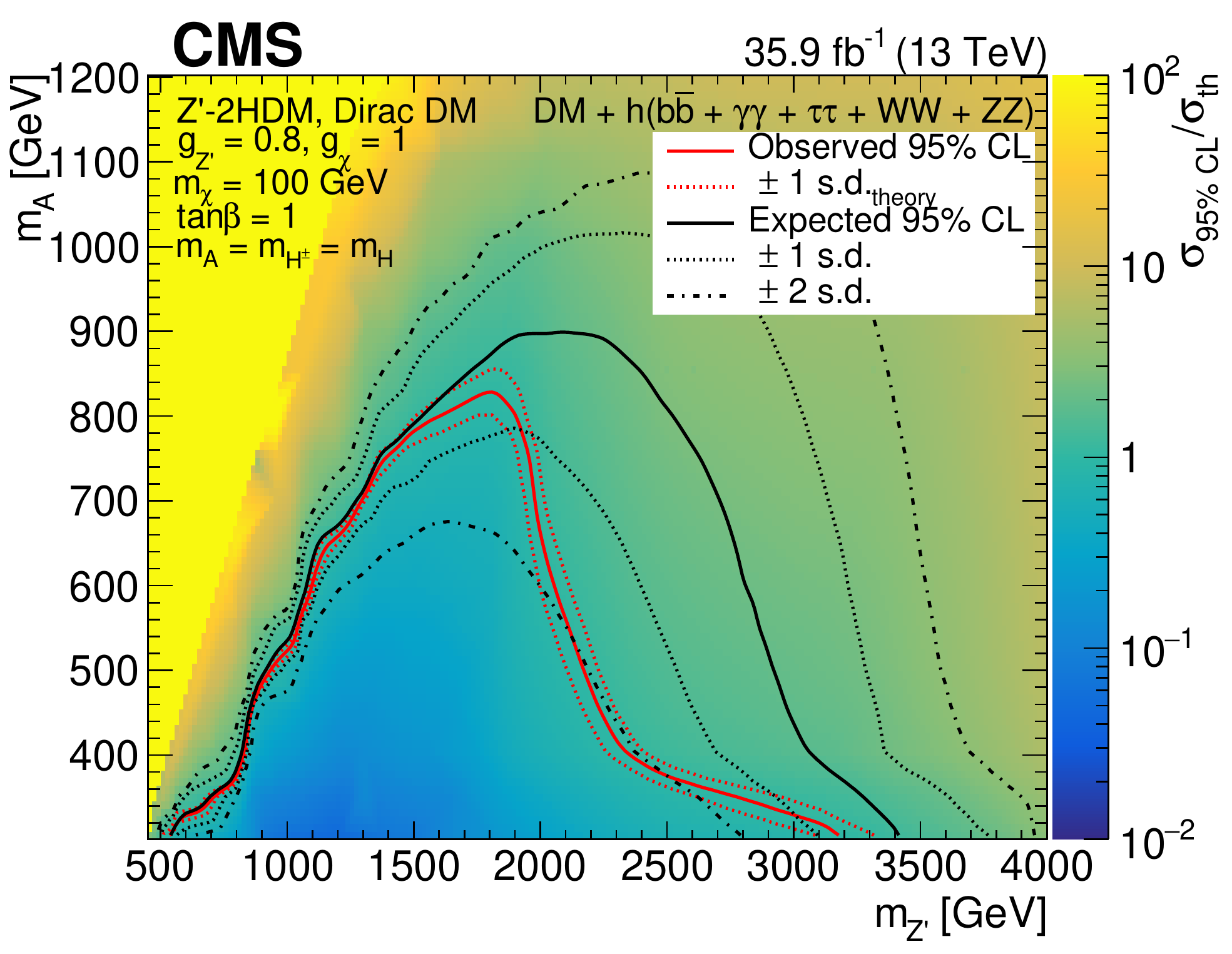}
    \includegraphics[width=0.495\textwidth]{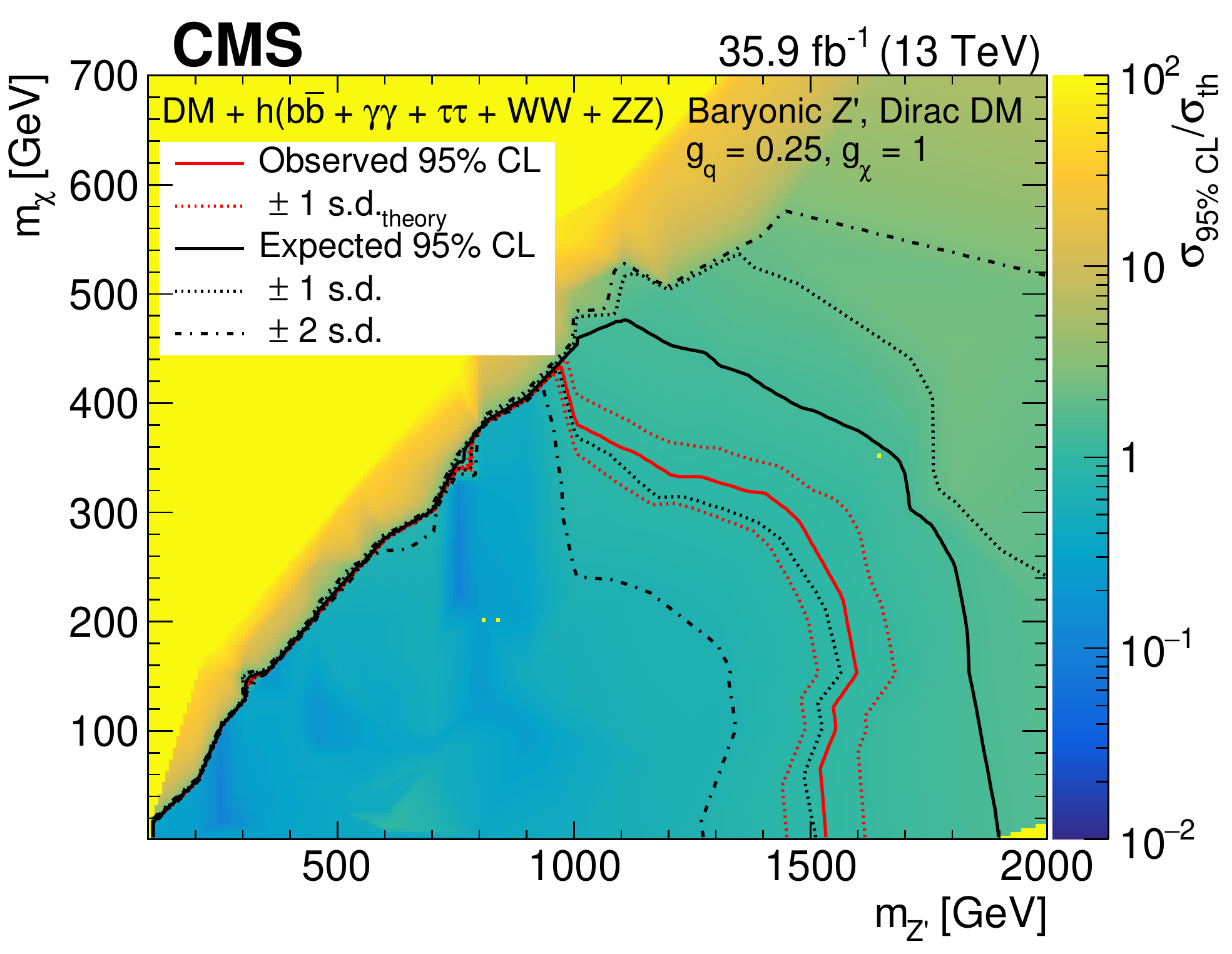}
    \caption{    The upper limits at 95\% \CL on the observed and expected $\sigma/\sigma_\text{th}$ in the \mZp--\mA and \mZp--\mdm planes for the \PZp-2HDM (left) and baryonic \PZp model (right), respectively. The region enclosed by the contours is excluded using the combination of the five decay channels of the Higgs boson for the following benchmark scenarios: $\gZp = 0.8$, $\gChi = 1$, $\tan\beta=1$, $\mdm = 100\GeV$, and $\mA = m_{\PH} = m_{\PH^\pm}$ for the \PZp-2HDM, and $\gChi = 1$,  $\gq = 0.25$ for the baryonic \PZp model.}
  \label{fig:monohcombo2d}
\end{figure}

Figure~\ref{fig:monohcombo2d} shows the observed and expected 95\% \CL exclusion contours on $\sigma/\sigma_\text{th}$ in the \mZp--\mA and \mZp--\mdm planes for the \PZp-2HDM (left) and baryonic \PZp (right) model, respectively.

\begin{figure}[htbp]
  \centering
    \includegraphics[width=0.475\textwidth]{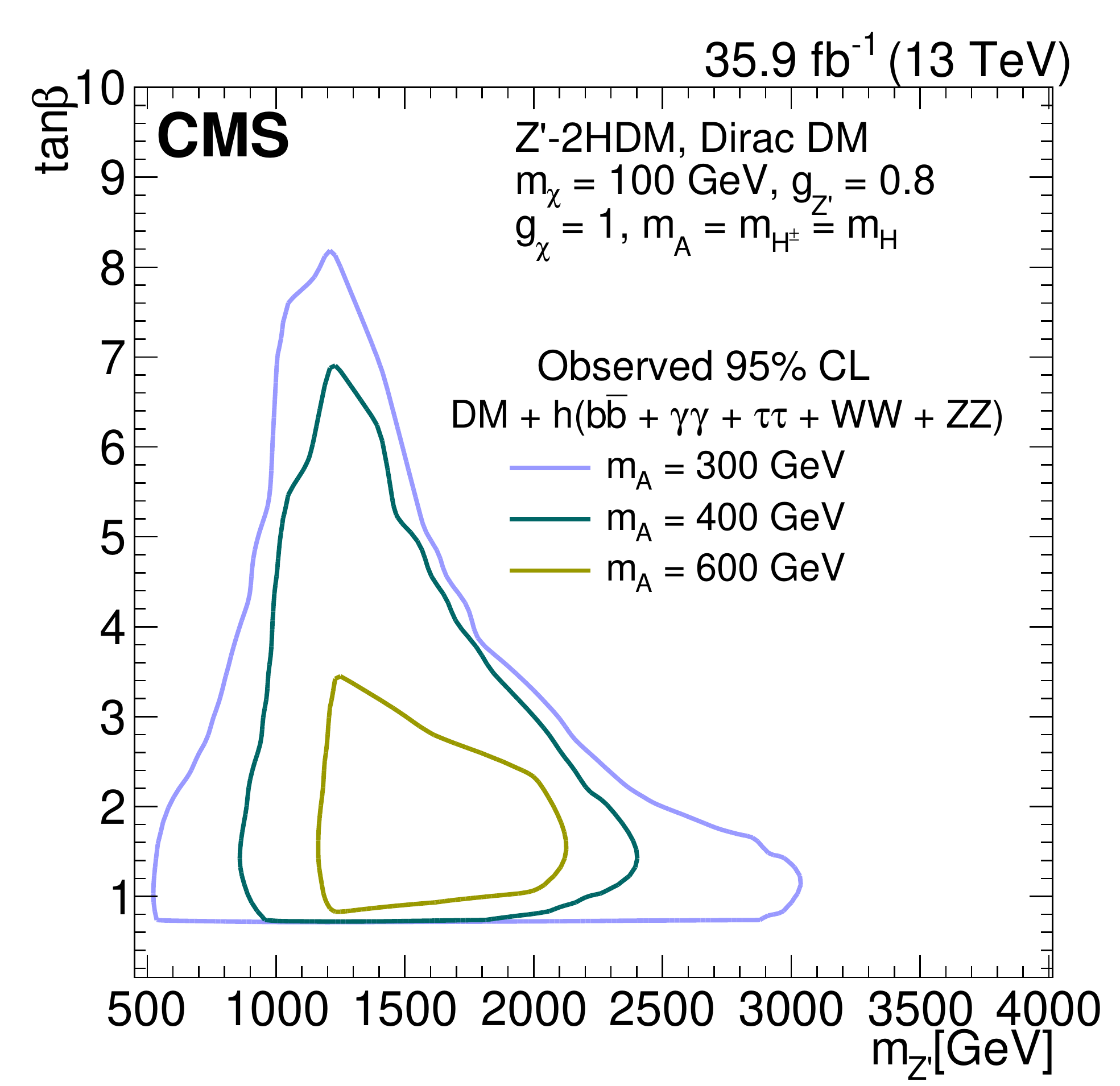}
  \caption{
  The upper limits at 95\% \CL on the observed $\sigma/\sigma_\text{th}$ for the \PZp-2HDM in the \mZp--$\tan\beta$ plane from the combination of the five Higgs boson decay channels.
  Each contour represents the excluded region for a given value of $\mA = 300$, 400, and 600\GeV.}
  \label{fig:monohcombotanbeta}
\end{figure}

The results for the \PZp-2HDM are also interpreted in the \mZp--$\tan\beta$ plane for three different \mA values: 300, 400, and 600\GeV. Since the shape of the \ptmiss distribution does not change with $\tan\beta$, and affects only the  product of the \PZp production cross section and branching fraction to the mono-\Ph channel, the limit shown in Fig.~\ref{fig:monohcombo2d} (left) can be simply rescaled for different values of $\tan\beta$, from 0.5 to 10. These limits, in the \mZp--$\tan\beta$ plane, are shown in Fig.~\ref{fig:monohcombotanbeta}. The area enclosed by the contour for a given value of \mA is excluded at 95\% \CL.

\begin{figure}[htbp]
  \centering
    \includegraphics[width=0.755\textwidth]{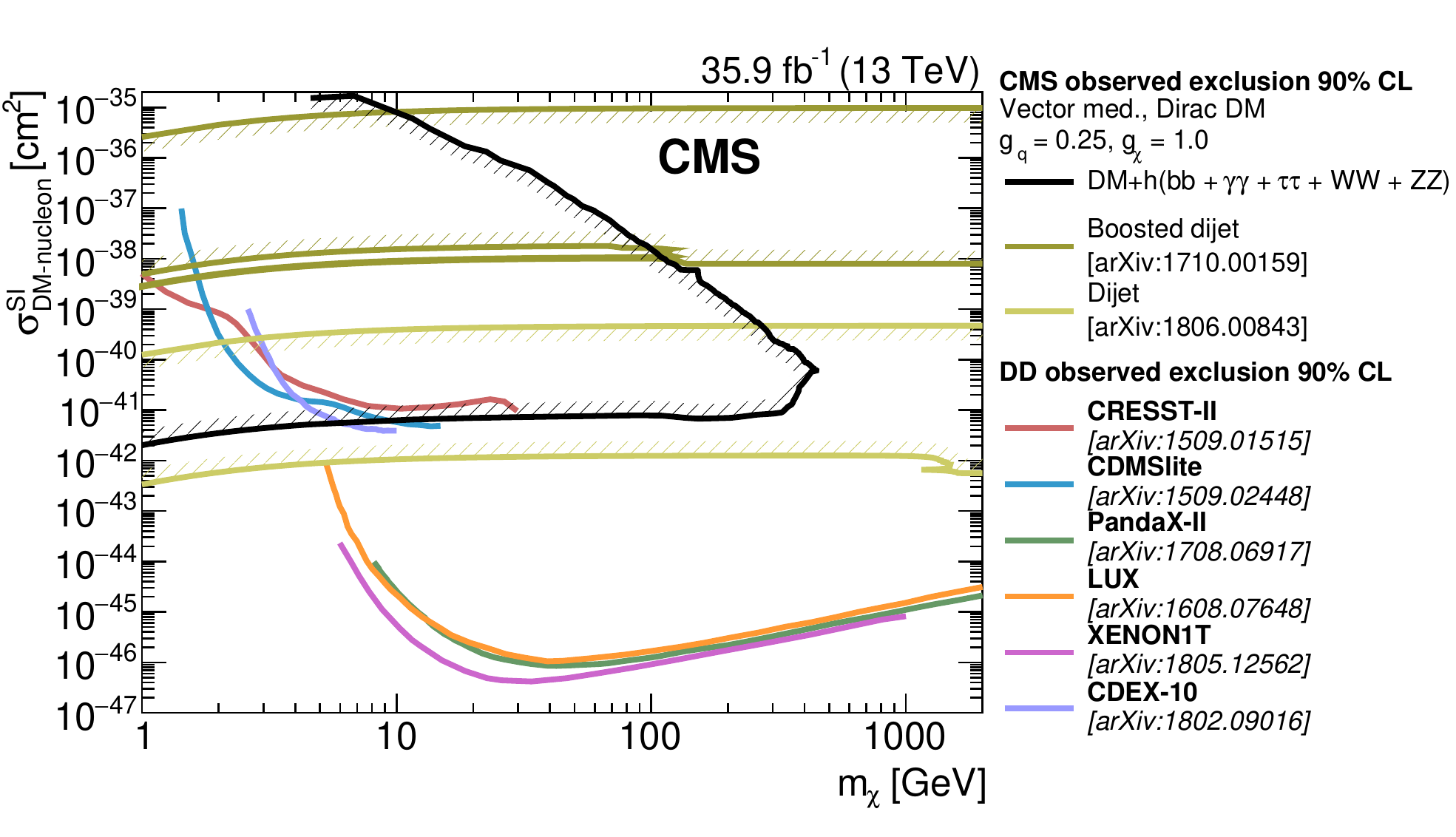}
  \caption{
The upper limits at 90\% \CL on the DM-nucleon spin-independent scattering cross section \SigSI, as a function of \mdm.
Results obtained in this analysis are compared with those from the CMS dijet analyses ~\protect\cite{Sirunyan:2017nvi,Sirunyan:2018xlo} and from several direct-detection experiments:
CRESST-II~\protect\cite{CresstII}, CDMSLite~\protect\cite{CDMSLite}, PandaX-II~\protect\cite{PandaxII}, LUX~\protect\cite{LUX}, XENON-1T~\protect\cite{XENON1T}, and CDEX-10~\protect\cite{CDEX10}. }
  \label{fig:monohcomboSI}
\end{figure}

Limits for the baryonic \PZp model are also interpreted in terms of limits on the $s$-channel simplified DM model proposed by the ATLAS-CMS Dark Matter Forum~\cite{Abercrombie:2015wmb} for comparison with direct-detection experiments. In this model, Dirac DM particles couple to a vector \PZp mediator, which also couples to the SM quarks. A point in the parameter space of this model is determined by four variables: the DM particle mass \mdm, the mediator mass \mMed, the mediator-DM coupling \gChi, and the universal mediator-quark coupling \gq. The couplings for the present  analysis are fixed to $\gChi = 1.0$ and $\gq = 0.25$, following the recommendation of Ref.~\cite{presentDM}. The results are interpreted in terms of  90\% \CL limits on the spin-independent (SI) cross section \SigSI for the DM-nucleon scattering. The value of \SigSI for a given set of parameters in the $s$-channel simplified DM model is given by~\cite{presentDM}:
\begin{equation}
	\SigSI = \frac{f^2(\gq)\gChi^2\mu^2_{\mathrm{nDM}}}{\pi \mMed^4},
\end{equation}
\noindent where $\mu_{\mathrm{nDM}}$ is the reduced mass of the DM-nucleon system and $f(\gq)$ is the mediator-nucleon coupling, which depends on \gq. The resulting \SigSI limits, as a function of
\mchi are shown in Fig.~\ref{fig:monohcomboSI}.
Results obtained in this analysis are compared with those from the CMS dijet \linebreak analyses \footnote[1]{We note that the limits presented in \cite{Sirunyan:2017nvi} are at 95\% CL, while the corresponding results at 90\% CL from that analysis are shown here.}  ~\protect\cite{Sirunyan:2017nvi,Sirunyan:2018xlo} and from several direct-detection experiments. For the chosen set of parameters, the cross section limit from the present analysis is more stringent than the direct-detection limits for \mchi between 1 and 5\GeV.

\section{Summary} \label{sec:summary}

A search for dark matter particles produced in association with a Higgs boson has been presented, using a sample of proton-proton collision data at a center-of-mass energy of 13\TeV, corresponding to an integrated luminosity of 35.9\fbinv. Results from five decay channels of the Higgs boson, $\Ph \to \bbbar$, \hgg, $\Ph \to \Pgt^{+}\Pgt^{-}$, $\Ph \to \PW^{+} \PW^{-}$, and \hzz, are described, along with their statistical combination. No significant deviation from the standard model prediction is observed in any of the channels or in their combination. Upper limits at 95\% confidence level on the production cross section of dark matter are set in a type-II two Higgs doublet model extended by a \PZp boson and in a baryonic \PZp model.
The results in the baryonic \PZp model are also interpreted in terms of the spin-independent dark matter nucleon scattering cross section.
This is the first search for DM particles produced in association with a Higgs boson decaying to a pair of $\PW$ or $\PZ$ bosons, and the first statistical combination based on five Higgs boson decay channels.

\begin{acknowledgments}
We congratulate our colleagues in the CERN accelerator departments for the excellent performance of the LHC and thank the technical and administrative staffs at CERN and at other CMS institutes for their contributions to the success of the CMS effort. In addition, we gratefully acknowledge the computing centers and personnel of the Worldwide LHC Computing Grid for delivering so effectively the computing infrastructure essential to our analyses. Finally, we acknowledge the enduring support for the construction and operation of the LHC and the CMS detector provided by the following funding agencies: BMBWF and FWF (Austria); FNRS and FWO (Belgium); CNPq, CAPES, FAPERJ, FAPERGS, and FAPESP (Brazil); MES (Bulgaria); CERN; CAS, MoST, and NSFC (China); COLCIENCIAS (Colombia); MSES and CSF (Croatia); RPF (Cyprus); SENESCYT (Ecuador); MoER, ERC IUT, PUT and ERDF (Estonia); Academy of Finland, MEC, and HIP (Finland); CEA and CNRS/IN2P3 (France); BMBF, DFG, and HGF (Germany); GSRT (Greece); NKFIA (Hungary); DAE and DST (India); IPM (Iran); SFI (Ireland); INFN (Italy); MSIP and NRF (Republic of Korea); MES (Latvia); LAS (Lithuania); MOE and UM (Malaysia); BUAP, CINVESTAV, CONACYT, LNS, SEP, and UASLP-FAI (Mexico); MOS (Montenegro); MBIE (New Zealand); PAEC (Pakistan); MSHE and NSC (Poland); FCT (Portugal); JINR (Dubna); MON, RosAtom, RAS, RFBR, and NRC KI (Russia); MESTD (Serbia); SEIDI, CPAN, PCTI, and FEDER (Spain); MOSTR (Sri Lanka); Swiss Funding Agencies (Switzerland); MST (Taipei); ThEPCenter, IPST, STAR, and NSTDA (Thailand); TUBITAK and TAEK (Turkey); NASU and SFFR (Ukraine); STFC (United Kingdom); DOE and NSF (USA).

\hyphenation{Rachada-pisek} Individuals have received support from the Marie-Curie program and the European Research Council and Horizon 2020 Grant, contract Nos.\ 675440, 752730, and 765710 (European Union); the Leventis Foundation; the A.P.\ Sloan Foundation; the Alexander von Humboldt Foundation; the Belgian Federal Science Policy Office; the Fonds pour la Formation \`a la Recherche dans l'Industrie et dans l'Agriculture (FRIA-Belgium); the Agentschap voor Innovatie door Wetenschap en Technologie (IWT-Belgium); the F.R.S.-FNRS and FWO (Belgium) under the ``Excellence of Science -- EOS" -- be.h project n.\ 30820817; the Beijing Municipal Science \& Technology Commission, No. Z181100004218003; the Ministry of Education, Youth and Sports (MEYS) of the Czech Republic; the Lend\"ulet (``Momentum") Program and the J\'anos Bolyai Research Scholarship of the Hungarian Academy of Sciences, the New National Excellence Program \'UNKP, the NKFIA research grants 123842, 123959, 124845, 124850, 125105, 128713, 128786, and 129058 (Hungary); the Council of Science and Industrial Research, India; the HOMING PLUS program of the Foundation for Polish Science, cofinanced from European Union, Regional Development Fund, the Mobility Plus program of the Ministry of Science and Higher Education, the National Science Center (Poland), contracts Harmonia 2014/14/M/ST2/00428, Opus 2014/13/B/ST2/02543, 2014/15/B/ST2/03998, and 2015/19/B/ST2/02861, Sonata-bis 2012/07/E/ST2/01406; the National Priorities Research Program by Qatar National Research Fund; the Ministry of Science and Education, grant no. 3.2989.2017 (Russia); the Programa Estatal de Fomento de la Investigaci{\'o}n Cient{\'i}fica y T{\'e}cnica de Excelencia Mar\'{\i}a de Maeztu, grant MDM-2015-0509 and the Programa Severo Ochoa del Principado de Asturias; the Thalis and Aristeia programs cofinanced by EU-ESF and the Greek NSRF; the Rachadapisek Sompot Fund for Postdoctoral Fellowship, Chulalongkorn University and the Chulalongkorn Academic into Its 2nd Century Project Advancement Project (Thailand); the Welch Foundation, contract C-1845; and the Weston Havens Foundation (USA). \end{acknowledgments}

\bibliography{auto_generated}
\cleardoublepage \appendix\section{The CMS Collaboration \label{app:collab}}\begin{sloppypar}\hyphenpenalty=5000\widowpenalty=500\clubpenalty=5000\vskip\cmsinstskip
\textbf{Yerevan Physics Institute, Yerevan, Armenia}\\*[0pt]
A.M.~Sirunyan$^{\textrm{\dag}}$, A.~Tumasyan
\vskip\cmsinstskip
\textbf{Institut f\"{u}r Hochenergiephysik, Wien, Austria}\\*[0pt]
W.~Adam, F.~Ambrogi, T.~Bergauer, J.~Brandstetter, M.~Dragicevic, J.~Er\"{o}, A.~Escalante~Del~Valle, M.~Flechl, R.~Fr\"{u}hwirth\cmsAuthorMark{1}, M.~Jeitler\cmsAuthorMark{1}, N.~Krammer, I.~Kr\"{a}tschmer, D.~Liko, T.~Madlener, I.~Mikulec, N.~Rad, J.~Schieck\cmsAuthorMark{1}, R.~Sch\"{o}fbeck, M.~Spanring, D.~Spitzbart, W.~Waltenberger, J.~Wittmann, C.-E.~Wulz\cmsAuthorMark{1}, M.~Zarucki
% ...
\vskip\cmsinstskip
\dag: Deceased\\
1:  Also at Vienna University of Technology, Vienna, Austria\\
% ...
\end{sloppypar}
\end{document}